\documentclass[12pt]{article}
\usepackage[left=1in,top=1in,right=1in,bottom=1in]{geometry}

\usepackage{geometry,setspace}

\usepackage{natbib}

\usepackage[hidelinks]{hyperref}

\usepackage{makeidx, siunitx, ragged2e}
\usepackage{booktabs,color}
\usepackage{amsfonts,bm}
\usepackage{amsmath}
\usepackage{amssymb}
\usepackage{graphicx}
\usepackage{rotating}
\usepackage{multirow}%
\usepackage{multicol}
\usepackage{subfigure}
\usepackage{rotating}
\usepackage{tikz}
\usepackage{pdfpages}
\usepackage{ulem}
\usepackage{booktabs}
\usepackage{lscape}
\usepackage{caption}
\usepackage{lipsum}
\usepackage{longtable}
\usepackage{booktabs}
\newcommand{\nc}{\normalcolor}

\begin{document}

\title{The Financial Market of Environmental Indices}

\author{
	Thisari K. Mahanama\thanks{Texas Tech University, Department of Mathematics \& Statistics, Lubbock, TX 79409-1042, USA, tmahanam@ttu.edu (Corresponding
		Author).}
	\and 
	Abootaleb Shirvani\thanks{Department of Mathematical Science, Kean University, Union, NJ, USA, ashirvan@kean.edu.}
	\and 
 	Svetlozar Rachev\thanks{Texas Tech University, Department of Mathematics
	\& Statistics, Lubbock, TX 79409-1042, USA, Zari.Rachev@ttu.edu.}
 \and 
 	Frank J. Fabozzi \thanks{Carey Business School, Johns Hopkins University, Baltimore, MD 21218, USA, fabozzi321@aol.com.}
}
\date{}
\maketitle

\begin{abstract}
This paper introduces the concept of a global financial market for environmental indices, addressing sustainability concerns and aiming to attract institutional investors. Risk mitigation measures are implemented to manage inherent risks associated with investments in this new financial market. We monetize the environmental indices using quantitative measures and construct country-specific environmental indices, enabling them to be viewed as dollar-denominated assets. Our primary goal is to encourage the active engagement of institutional investors in portfolio analysis and trading within this emerging financial market.

To evaluate and manage investment risks, our approach incorporates financial econometric theory and dynamic asset pricing tools. We provide an econometric analysis that reveals the relationships between environmental and economic indicators in this market. Additionally, we derive financial put options as insurance instruments that can be employed to manage investment risks. Our factor analysis identifies key drivers in the global financial market for environmental indices.

To further evaluate the market's performance, we employ pricing options, efficient frontier analysis, and regression analysis. These tools help us assess the efficiency and effectiveness of the market. Overall, our research contributes to the understanding and development of the global financial market for environmental indices.
\\

 
\end{abstract}


\newpage

\section{Introduction}\label{sec: Introduction}


\textbf{Introduction:}

This paper investigates a novel approach to developing a financial market of environmental indices. This approach is an alternative to the existing financial markets of weather derivatives and catastrophe (CAT) bonds.

Rather than delving into the socioeconomic factors influencing a country’s environment, the paper aims to gather information from the existing literature on these factors to create a financial index. This index can be traded as a financial asset with a price process aligned with financial econometrics and dynamic asset pricing. The index will be a financial asset within the financial markets of environmental assets that will be developed in this paper.

The purpose of this literature review is to propose a method for creating a Global Dollar Environmental Index (GDEI) based on economic factors influencing the environment. This index can be used to evaluate the environmental impact of various countries worldwide. The absence of such an index makes it challenging to quantify financial losses due to adverse environmental impacts. The need for a global environmental index arises from the necessity of measuring the environmental consequences of economic activities and developing policies to mitigate these negative impacts.\\

\textbf{Problem statement:}

There are concerns about the environmental ramifications of economic development and the lack of a financial index to assess the financial losses and economic factors affecting the environment. However, accurately measuring the environmental impact of different countries using an index is a complex task. Concern over the environmental consequences of economic development has highlighted the need for a new environmental index that can effectively assess the financial losses and economic factors impacting the environment. While existing indices like the Environmental Performance Index (EPI), Ecological Footprint Index (EFI), Green GDP, and Environmental Sustainability Index (ESI) have made valuable contributions, they primarily focus on evaluating the state of the environment rather than specifically addressing the economic activities that drive environmental impacts.

The Environmental Performance Index (EPI)\href{https://epi.yale.edu/}{} is a widely recognized index that provides a comprehensive measure of a country's overall environmental sustainability. It evaluates various indicators related to environmental health, ecosystem vitality, social well-being, governance, and economic strength. However, its emphasis is on assessing the state of the environment rather than explicitly considering the economic activities that contribute to environmental degradation or preservation.

The Ecological Footprint Index (EFI)\href{https://www.footprint}{} serves as a valuable tool for quantifying the environmental impact of human activities. It assesses the amount of natural resources and the ecological services required to sustain these activities, providing an estimate of the resources needed to support a specific population's consumption patterns and waste generation. While the EFI helps determine the strain placed on the environment, it does not directly evaluate the economic activities driving these ecological footprints.

The Green GDP\href{https://en.wikipedia.org/wiki/Green_gross_domestic_product}{}\footnote{Green GDP = GDP – Environmental Costs – Social Costs.} is an economic indicator that adjusts the conventional gross domestic product (GDP) by factoring in the value of environmental resources and the costs associated with environmental degradation. By incorporating these environmental impacts, the Green GDP aims to provide a more comprehensive measure of economic performance that takes sustainability into account. However, it does not specifically focus on evaluating the environmental state or the effectiveness of environmental policies.

The Environmental Sustainability Index (ESI)\href{https://link.springer.com/referenceworkentry/10.1007/978-3-642-28036-8_116#:~:text=Definition,sustainability%20at%20a%20national%20level.}{}\footnote{The Environmental Sustainability Index (ESI) measures the overall progress of nations toward environmental sustainability. As a composite index, it tracks a set of environmental, socioeconomic, and institutional indicators that characterize and influence environmental sustainability at a national level.} assesses a country's overall environmental sustainability based on various indicators related to environmental health, ecosystem vitality, social and institutional capacity, and global stewardship. While the ESI provides a comprehensive measure of a country's performance in managing environmental challenges, it does not directly capture the economic activities driving environmental impacts or financial losses.

Therefore, there is a need for a new environmental index that addresses the limitations of existing indices and specifically focuses on evaluating the economic activities affecting the environment. This new index should consider both the state of the environment and the economic factors contributing to environmental degradation or preservation. By incorporating indicators that assess the impact of economic development on the environment, such as resource consumption, pollution levels, and the costs associated with environmental damage, this new index would provide a more comprehensive understanding of the environmental consequences of economic development. It would enable policymakers and researchers to make informed decisions that promote sustainable development while also mitigating the adverse effects of economic activities on the environment.\\

\textbf{Proposed approach:}

Our proposal entails the establishment of a global financial market dedicated specifically to environmental indices across various countries worldwide. This market would provide a comprehensive and worldwide perspective on each country's environmental impact and interdependencies. The increasing importance of sustainability for investors emphasizes the need for a financial market that focuses on environmental indices. However, like any investment, there are inherent risks, necessitating risk mitigation measures in the financial industry.

The primary objective of our proposal is to establish a financial market for environmental indices, treating them as risky financial assets. To address this challenge, we suggest a unique approach that involves assessing and monetizing environmental indices through quantitative measures of environmental performance derived from the World Bank's world development indicators. Our methodology involves constructing an environmental index for each country that combines equally weighted economic and environmental factors. By multiplying each country's index by the GDP per capita, we assign a monetary value to these indices. This enables us to conduct financial econometric forecasting and predict the future behavior of environmental indices as financial assets.

Obtaining dollar-denominated environmental indices for each country allows us to create a market consisting of risky financial assets. This global dollar-denominated index resembles an equally weighted market index, with each country's environmental index representing a risky asset within the overall market index.

To generate efficient frontiers, we employ traditional optimization analysis using these dollar-denominated environmental indices from major developed countries worldwide. We anticipate active engagement from institutional investors in trading these assets, similar to other derivative products. Additionally, we utilize portfolio theory to analyze the financial market of these environmental indices. Through various risk-return measures, we assess the risk and return characteristics of the indices. This analysis helps determine the optimal allocation of investments among the indices, considering their respective risk profiles and potential returns.

Our approach incorporates financial economic theory into the evaluation and management of risks stemming from adverse movements in both environmental and economic indices. We utilize methods developed for financial portfolio insurance and adopt a comprehensive perspective on a country's environmental conditions using dynamic asset pricing theory tools. Within this framework, we consider environmental, social, and governance (ESG)\footnote{Environmental, social, and corporate governance (ESG), also known as environmental, social, governance, is a business framework for considering environmental issues and social issues in the context of
corporate governance.} factors as a subset of environmental and economic factors that significantly influence a country's environmental conditions.

To compare the environmental indices of different countries, we employ econometric analysis and dynamic asset pricing approaches. This analysis provides insight into the relationships between environmental and economic indicators, shedding light on their impact on the financial value of the indices. Portfolio-efficient frontier techniques are applied, accounting for different risk measures, to effectively evaluate the indices.

As part of our financial market analysis, we derive financial put options that serve as insurance instruments for the environmental indices. These put options assist the financial industry in assessing and managing potential risks associated with adverse movements in the indices.

Factor analysis is commonly employed to identify influential factors that impact asset portfolio returns, including stocks, bonds, and other financial instruments. Our factor analysis focuses on the GDEI, utilizing each country's index as a factor. This enables the identification of key drivers behind environmental degradation and economic growth, fostering the formulation of sustainable economic development policies. Additionally, this analysis reveals interrelationships between countries and their association with the overall GDEI score. Examining factor loadings enhances our understanding of risk and return sources within economic and environmental indicator portfolios.
\\
\\

\textbf{Challenges in creating the market for the global environmental index:}

When examining the market in each individual country, we observe the beta value for the capital asset pricing model (CAPM)\href{https://www.investopedia.com/terms/c/capm.asp}{}\footnote{The capital asset pricing model (CAPM) describes the relationship between the systematic risk, or the general perils of investing, and the expected return for assets, particularly stocks. }. We aim to establish the foundation of portfolio insurance for the first time. Our plan involves introducing derivatives based on the global environmental index, followed by the creation of a CAPM and  Jensen’s alpha index\footnote{Jensen's measure, or Jensen's alpha, is a risk-adjusted performance measure that represents the average return on a portfolio or investment, above or below that predicted by the capital asset pricing model (CAPM), given the portfolio's or investment's beta and the average market return.}\href{https://www.investopedia.com/terms/j/jensensmeasure.asp#:~:text=The%20Jensen's%20measure%2C%20or%20Jensen's,and%20the%20average%20market%20return.}. 
 Portfolio insurance\footnote{Portfolio insurance is the strategy of hedging a portfolio of stocks against market risk by short-selling stock index futures.} \href{https://www.investopedia.com/terms/p/portfolioinsurance.asp#:~:text=Portfolio%20insurance%20is%20a%20hedging,also%20refer%20to%20brokerage%20insurance.}{} is a comparable concept in this context. Consequently, industries will purchase puts on the indices of individual countries if they have invested in the index and if certain countries are facing environmental challenges.

To obtain Jensen’s alpha, we conducted a robust regression analysis. According to the CAPM, alpha should be equal to 0, but in our market, alpha deviates from 0 due to non-equilibrium environmental conditions.

Jensen's alpha represents a one-factor model. We perform a regression of an index, such as the US market index, against the entire world, which constitutes a one-factor model. Additionally, we plan to develop analogs of the Fama-French three-factor and five-factor models.

Our objective is to establish a financial market for environmental indices, with the global index serving as the central component. Each individual country functions as an asset within this global index, similar to the Dow Jones 30 stocks and those in the Dow Jones Industrial Average (DJIA)\footnote{The Dow Jones Industrial Average (DJIA) is a stock market index that tracks 30 large, publicly owned blue-chip companies trading on the New York Stock Exchange (NYSE) and Nasdaq.}\href{https://www.investopedia.com/terms/d/djia.asp#:~:text=The%20Dow%20Jones%20Industrial%20Average%20(DJIA)%20is%20a%20stock%20market,his%20business%20partner%2C%20Edward%20Jones.}{}.
This is not a traded asset like the DJIA; it is more like an exchange-traded fund (ETF) replicating the DJIA\footnote{For investors optimistic that US stocks will keep rebounding from their recent lows, one Dow Jones Industrial Average ETF provides focused exposure to the market.} \href{https://www.investopedia.com/etfs/etfs-track-dow/}{} or SPDR (SPY), which is an ETF replicating the S\&P 500 index.

The existing literature indicates that current environmental indices lack a crucial component: a dynamic time series econometric model capable of accurately forecasting price movements. Consequently, these indices fail to provide a means for traders to incorporate them into their financial portfolios effectively. To our knowledge, no research has been conducted on financial instruments specifically designed to assess a country's environmental downturns. In response to this gap, we aim to establish a financial market for environmental indices. This market would enable the financial industry to actively engage with and manage potential future downturns in a country's environmental conditions. While the significance of ESG factors in financial markets is acknowledged when it comes to addressing environmental concerns, our objective transcends this focus, encompassing the overall environmental state of a country.\\

The structure of this paper is as follows: First, we establish dollar-denominated environmental indices for 10 countries, utilizing the 14 world development indicators introduced in Section 2. Subsequently, we present a global US-dollar-denominated index of the environment and examine the dynamics of asset prices. In Sections 3 and 4, we assess the impact of adverse events in the individual country indices on the global environmental index. We then investigate the financial market for these indices, employing the tools of dynamic asset pricing theory. To accomplish this, we offer the optimal portfolio weight composition and an efficient frontier in Section 5. In Section 6, we propose an option pricing model for the global environmental index. Additionally, in Section 7, we conduct factor analysis on the GDEI by considering each country index as a factor. Finally, our concluding remarks are presented in Section 8.

\section{World Development Economic and Environmental Indicators} \label{sec: DataDescription}
\
\
\
Environmental indicators play a crucial role in assessing the environmental impact and ecological sustainability of a nation. These indicators include the CO\textsubscript2 emissions, nitrous oxide emissions, methane emissions, forest area, agricultural land, arable land, land area, and surface area. Carbon dioxide emissions provide insight into a country's carbon footprint and its contribution to climate change. Nitrous oxide and methane emissions indicate the extent of a country's involvement in other significant greenhouse gas emissions. The forest area measures the extent of the forest cover, which is important for biodiversity and carbon sequestration. The agricultural land and arable land indicators highlight a country's agricultural potential and food security. The land area and surface area provide context for analyzing the scale and geographical characteristics of a country's resources.

Economic indicators are equally essential for evaluating the economic sustainability and development of a nation. These indicators include the GDP per capita, adjusted net savings, energy intensity level of primary energy, renewable energy consumption, and total natural resources rents. The GDP per capita reflects the economic well-being of a country's population. Adjusted net savings measure a nation's capacity to maintain and enhance its economic resources over time. The energy intensity level of primary energy indicates the efficiency of a country's energy consumption. The renewable energy consumption assesses a country's progress towards a sustainable energy transition. The total natural resources rents capture the economic benefits derived from natural resources, accounting for both renewable and non-renewable resources.

Descriptions of the aforementioned
world development indicators \citep{Data} are listed in the Appendix.

\section{The Financial Environmental Market}
\subsection{Constructing a global dollar environmental index}
\
\
\
Within this section, we develop a comprehensive worldwide environmental index by employing the 14 world development indicators, which are described in detail in the Appendix. We focus on 10 countries with the world's largest economies: the United States, China, Japan, Germany, the United Kingdom, India, Brazil, Australia, France, and Canada. To ensure accuracy, we rely on data provided in the data in (IBRD, 2022),
specifically utilizing reported data spanning from 2007 to 2021.

\begin{figure}
            \centering
            \subfigure{\includegraphics[width=0.42\textwidth]{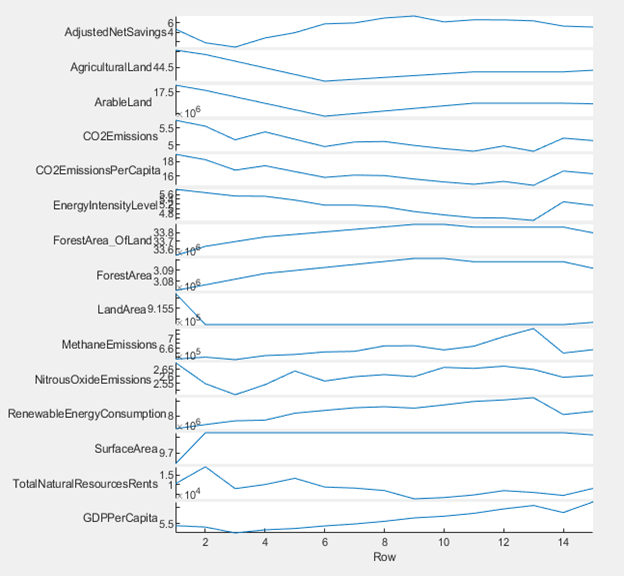}}
             \subfigure[]{\includegraphics[width=0.42\textwidth]{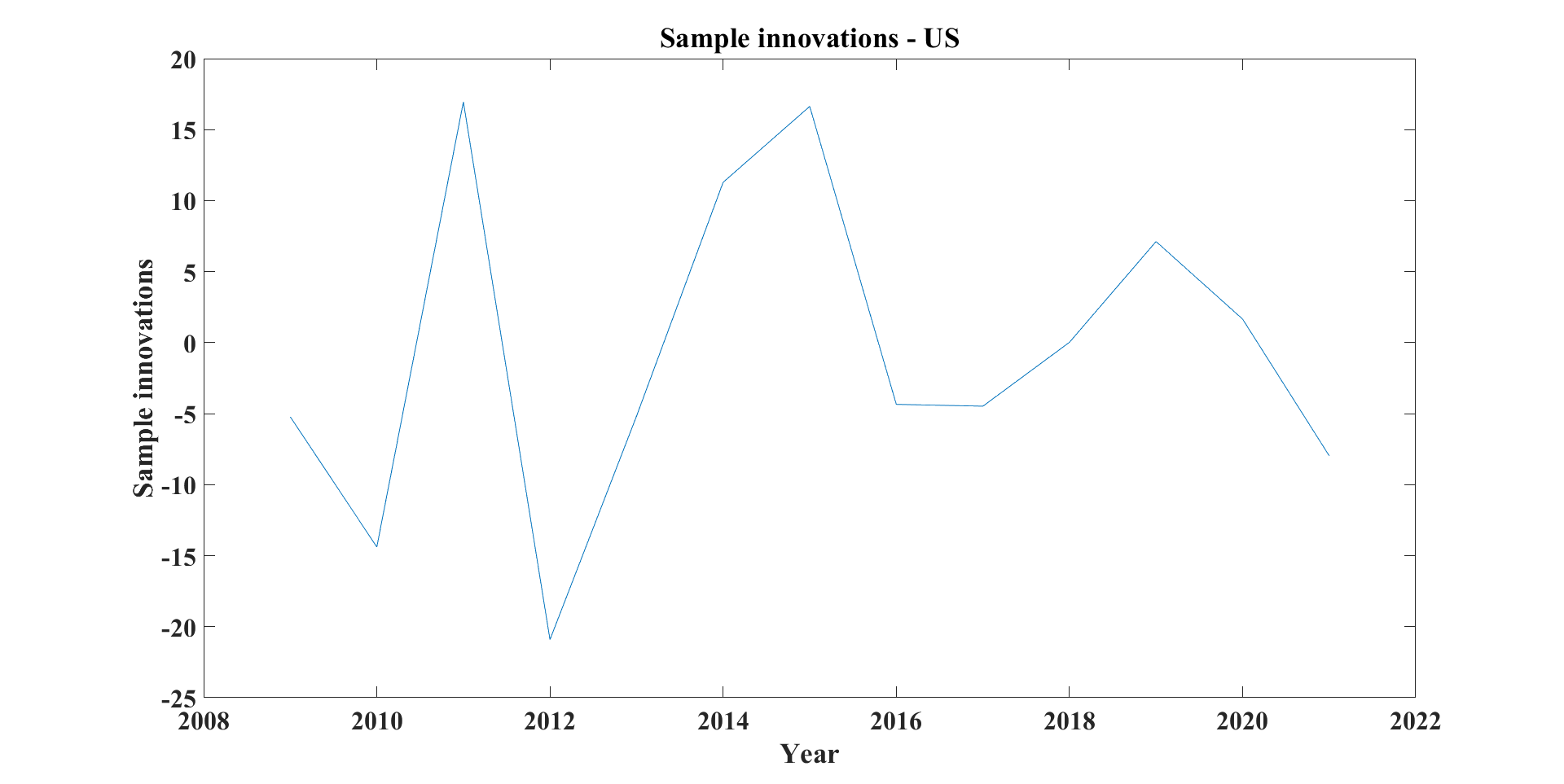}}
           
            \caption{(a) The US development indicators according to the World Bank reports \citep{Data} from 2007 to 2021 and b) the log returns of the exponentially transformed Dollar Environmental Index (DEI) subject to the constraints in Eq. (\ref{Eq:f(x)})}
            \label{Fig-US-indicators}
        \end{figure}

We denote by $F(k,l)$ the $k$\textsuperscript{th} world development indicator ($k=1,\cdots,K=14$) for the $l$\textsuperscript{th} country ($l=1,\cdots,L=10$) for a given year,
such that all indicators are strictly positive, i.e., $F(k,l)>0$ for all $k,l$,
so that they positively contribute to the environmental index.
The US development indicators that positively contribute to the US environmental index are shown in Figure \ref{Fig-US-indicators}(a).
        
\subsubsection{Standardized indicator}
\
\
\
We normalize each indicator of a country, $F(k,l)$, using the corresponding indicator of all the countries ($F(k,l),  \;\; l=1,\cdots,L=10$) as follows:
\begin{equation} \label{Eq:FN(k,l)}
F\textsubscript N(k,l)=\frac{F(k,l)}{\sum_{l=1}^{L} F(k,l)}, \;\;\;\; k=1,\cdots, K,  \;\;\; l=1,\cdots,L.
\end{equation}

\subsubsection{Environmental index for a country}

\
\
\
We then define the environmental index (EI) for country $l$, $EI(l)$, as the average of its normalized indicators, excluding the GDP:
\begin{equation}
\label{Eq:EI(l)}
EI(l)=\frac{1}{K} \sum_{i=1}^{K} F{\textsubscript{N}}(k,l),   \,\,\,\, l=1,\cdots,L,
\end{equation}
\\
so that $EI(l) \in (0,1) , \; l=1,\cdots,L$. 

We monetize the US dollar value of $EI(l)$ for year $t$, $WI_t(l)$, by weighting it with its corresponding GDP per capita, $GDP t(l)$, to define a US-dollar-denominated environmental index, the Dollar Environmental Index (DEI), for country $l$ for year $t$ as follows:

\begin{equation} \label{Eq:DEI(l)}
DEI_t(l)=GDP_t(l) \cdot WI_t(l),  \;\;\; t=t_0=2007,\cdots,t_{15}=2021. 
\end{equation}
This indicates the ``environment" of a country $l$, denominated in US dollars.

\subsubsection{Global dollar environmental index}

\
\
\
To compare the environment of a country with that of the whole world, we construct a global DEI by taking the average of the DEIs:
\begin{equation} \label{Eq:DEI}
GDEI_t=\frac{1}{L} \sum_{l=1}^{L} DEI_t(l),  \;\;\; t=t_0,\cdots,t_{15}.
\end{equation}




\subsection{Financial and econometric modeling of environmental indices }
\
\
\
To model the environmental index within the framework of a financial market, we can adapt the dynamic asset pricing theory approach. This approach involves applying principles and techniques used in the dynamic asset pricing theory to analyze and understand the dynamics of the environmental index. 
One way to do this is to employ exponential transformations of the time series data for the environmental index \citep{schoutens2003levy, duffie2010dynamic}. The exponential transformation helps mimic the behavior of asset prices in financial markets, allowing us to apply similar modeling techniques. This transformation enables us to analyze the volatility  risk and potential financial losses associated with the big downturns of the environmental indices. 
We determine the exponential transformation for each DEI, considering the years $t=t_0,\cdots,2021$, as follows:
\begin{equation} \label{Eq:f(x)}
\begin{split}
f(x) &= a\exp(bx), \;\;\;\; a>0, b>0,\\
f\left(\min_{l,t} \;\; DEI_t(l)\right)&=0, \\
f\left(\max_{l,t} DEI_t(l)\right)&=1. \\
\end{split}
\end{equation}	

Under this optimization, we set the exponential transformation of the lowest DEI to 0 and that of the highest DEI to 1.
We assume that the ``asset price" should have a minimum value of 0 and a maximal value of 1, representing 100\%. This scale is used in ESG rankings \citep{scatigna2021achievements}.
The exponential transformation for the US DEI produces $a=0.0037$ and $b=0.0001$. 
The exponential transformation is necessary, as we will define the CAPM and compute the CAPM beta value for each  country with respect to the global index.
With all DEIs positive, we define the log returns for the exponentially transformed DEI and thus introduce the environment asset pricing model for each country with respect to the global index. 

To obtain a stationary series, we apply the log return transformation to the exponentially transformed DEIs, as is commonly done in dynamic asset pricing theory \citep{duffie2010dynamic}. Specifically, for each country $l=1,\cdots,10$ and the global index $l=10$, we define the log return of the exponentially transformed DEI at time $t$, denoted by $R_t(l)$, as follows:
\begin{equation} \label{Eq:logReturns}
R_t(l) = \log \frac{f(DEI_t(l))}{f(DEI_{t-1}(l))}, \;\;\;\; t=t_0, \cdots ,t_{15},
\end{equation}
\
\
\
where $f(DEI_t(l))$ is the exponentially transformed DEI for the $l$\textsuperscript{th} country for year $t$.
\label{sec:TS}
Next, for each country $l=1,\cdots,10$, we model the log returns $R_t(l)$ using an autoregressive AR(1) process:
\begin{equation} \label{Eq:R_t(l)}
R_t(l) = \phi_{0} + z_{t} + \theta_{1} z_{t-1}, \;\;\;  t=t_0, \cdots ,t_{15},
\end{equation}
\
\
\
where $\phi_{0}$ and $\theta_{1}$ are parameters to be estimated, and $z_{t}$ is a Gaussian innovation process with a mean of zero and a variance $\sigma_{t}^2$. Specifically, $z_{t}$ is defined as $z_{t}=\sigma_{t}\epsilon_{t}$, where the $\epsilon_{t}$ are assumed to be independent and identically distributed (iid) Gaussian random variables. This process allows the modeling of the dependence structure of the log returns, which is important for capturing the dynamics of the DEIs over time.

To model the volatility $\sigma_{t}$ of the log returns, we consider three popular time-varying volatility models: the autoregressive conditional heteroskedasticity (ARCH) model, the generalized autoregressive conditional heteroskedasticity (GARCH) model, and the exponential general autoregressive conditional heteroskedastic (EGARCH) model. The ARCH model is used to estimate risk by providing a model of volatility that more closely resembles real markets. The GARCH model assumes that the variance of the error term follows an autoregressive moving-average process, and the EGARCH model extends the ARCH and GARCH models by allowing asymmetric effects of positive and negative shocks on volatility, which are relevant in financial markets. We estimate the parameters of each model and select the model that yields the best fit to the data.

In practice, we estimate the parameters of each model using maximum likelihood estimation, and we choose the model that minimizes a goodness-of-fit criterion, such as the Akaike information criterion (AIC) or the Bayesian information criterion (BIC). The resulting estimates of the parameters can be used to forecast the volatility of the time series, which is important for risk management and portfolio optimization purposes. The ARCH, GARCH, and EGARCH models are widely used in financial econometrics and have been shown to provide accurate forecasts of volatility in a variety of financial time series data.\footnote{For more information about these models, see \cite{tsay2005analysis} and \cite{hamilton2020time}.}

The GARCH(1,1) model is one of the most widely used time-varying volatility models and is defined as follows:
\begin{equation} \label{Eq:GARCH}
\begin{split}
    \sigma_{t} &= \frac{z_{t}}{\epsilon_{t}}, \\
    \sigma_{t}^{2} &= \alpha_{0}+\alpha_{1}z_{t-1}^{2}+\beta_{1}\sigma_{t-1}^{2}, \;\;\; t=t_0, \cdots ,t_{30},
\end{split}
\end{equation}
\
\
\
where $\alpha_{0}$, $\alpha_{1}$, and $\beta_{1}$ are parameters to be estimated, and $z_{t}$ is the innovation process from the AR(1) model defined in Eq. (\ref{Eq:R_t(l)}). The sample innovations $\epsilon_{t}$ are assumed to be iid Gaussian random variables with zero mean and unit variance \citep{tsay2005analysis}. The GARCH(1,1) model captures the volatility clustering phenomenon, where large returns tend to be followed by large returns and vice versa, and it is well-suited to modeling financial time series data \citep{bollerslev1986generalized}.

In practice, we estimate the parameters of the GARCH(1,1) model using the maximum likelihood estimation method, which is a commonly used statistical technique for estimating the parameters of a model given some observed data. We then select the model that minimizes a goodness-of-fit criterion, such as the AIC or BIC, which are widely used in statistical inference and model selection. The resulting estimates of the parameters can be used to forecast the volatility of the log returns, which is important for risk management and portfolio optimization purposes.

To model the log returns $R_{t}(l)$ in Eq. (\ref{Eq:R_t(l)}), we consider three univariate models with standard normal iid innovations: Model 1 (AR(1)-ARCH(1)), Model 2 (AR(1)-GARCH(1,1)), and Model 3 (AR(1)-EGARCH(1,1)). These models are commonly used in financial time series analysis and have been shown to provide accurate forecasts of volatility.

To select the best model for each country, we compare the performances of the three models based on the AIC and BIC, which are commonly used in statistical inference and model selection. Table \ref{AIC_BIC} presents the AIC and BIC values for each model and country. For each country, we select the model that results in the lowest AIC and BIC values among the estimated models (Models 1, 2, and 3). For example, we find that Model 1 outperforms the other models in modeling the log returns of the US DEI.

To simulate future log returns for the 10 countries, we combine the simulated sample innovations from each country to obtain a 10-dimensional sample of 30 observations. We employ $S=10,000$ scenarios for the innovations based on a fitted 10-dimensional normal-inverse Gaussian distribution (NIG), which is in the domain of attraction of a 10-dimensional multivariate Gaussian distribution \citep{oigaard2004multivariate}. By using NIG innovations instead of normal innovations, we preserve the asymptotic unbiasedness of the parameters in the marginal time series models (Eq. (\ref{Eq:GARCH})).

To generate Monte Carlo scenarios of the innovations for the log returns in the year 2022, $R_{t_{31}}(1;s),....,R_{t_{31}}(10;s)), s=1,...,S$, we use the estimated parameters in Models 1, 2, and 3 for each of the 10 marginal (one-dimensional) time series. We generate $S=10,000$ scenarios of the innovations, as required by the Basel II accord for properly assessing the tail risk in the portfolio of returns \citep{jacobson2005credit, orgeldinger2006basel}. It is worth noting that to assess the confidence bound for these parameters, bootstrapping methods would be required, but this is beyond the scope of this paper. The accurate simulation of future log returns is important for risk management and portfolio optimization purposes, as it enables investors to estimate the potential losses and gains of their investments and adjust their portfolios accordingly.
Using the Monte Carlo simulation approach described above, we obtain $S$ scenarios for the log returns of the 10 environmental indices for the year 2021. The use of NIG innovations allows us to capture the tail dependencies of the indices, which are important in modeling financial markets. As a result, our overall forecast of the environmental economic market for 2022 exhibits all the ``stylized facts" of a financial market, as documented in the literature \citep{rachev2000stable, taylor2011asset}. These stylized facts include, among others, fat tails, volatility clustering, and the non-normality of the log returns. 
\begin{table}[h]
\centering
\begin{tabular}{@{}l|ccc|ccc|@{}}
\cmidrule(l){2-7}
                                       & \multicolumn{3}{c|}{\textbf{AIC}}                                                                & \multicolumn{3}{c|}{\textbf{BIC}}                                                                \\ \cmidrule(l){2-7} 
                                       & \multicolumn{1}{c|}{\textbf{Model 1}} & \multicolumn{1}{c|}{\textbf{Model 2}} & \textbf{Model 3} & \multicolumn{1}{c|}{\textbf{Model 1}} & \multicolumn{1}{c|}{\textbf{Model 2}} & \textbf{Model 3} \\ \midrule
\multicolumn{1}{|l|}{\textbf{US}}      & \multicolumn{1}{c|}{18.959}           & \multicolumn{1}{c|}{20.959}           & 21.29            & \multicolumn{1}{c|}{21.219}           & \multicolumn{1}{c|}{23.784}           & 24.114           \\ \midrule
\multicolumn{1}{|l|}{\textbf{Brazil}}  & \multicolumn{1}{c|}{29.362}           & \multicolumn{1}{c|}{31.362}           & 25.261           & \multicolumn{1}{c|}{31.622}           & \multicolumn{1}{c|}{34.187}           & 28.085           \\ \midrule
\multicolumn{1}{|l|}{\textbf{Germany}} & \multicolumn{1}{c|}{30.849}           & \multicolumn{1}{c|}{31.803}           & 23.499           & \multicolumn{1}{c|}{33.109}           & \multicolumn{1}{c|}{34.628}           & 26.323           \\ \midrule
\multicolumn{1}{|l|}{\textbf{India}}   & \multicolumn{1}{c|}{15.352}           & \multicolumn{1}{c|}{16.545}           & 6.4929           & \multicolumn{1}{c|}{17.611}           & \multicolumn{1}{c|}{19.37}            & 9.3176           \\ \midrule
\multicolumn{1}{|l|}{\textbf{UK}}      & \multicolumn{1}{c|}{44.244}           & \multicolumn{1}{c|}{46.244}           & 39.406           & \multicolumn{1}{c|}{46.504}           & \multicolumn{1}{c|}{49.069}           & 42.23            \\ \midrule
\multicolumn{1}{|l|}{\textbf{Canada}}      & \multicolumn{1}{c|}{44.724}           & \multicolumn{1}{c|}{46.718}           & 43.304           & \multicolumn{1}{c|}{46.984}           & \multicolumn{1}{c|}{49.543}           & 46.129           \\ \midrule
\multicolumn{1}{|l|}{\textbf{France}}      & \multicolumn{1}{c|}{36.781}           & \multicolumn{1}{c|}{38.101}           & 36.897           & \multicolumn{1}{c|}{39.041}           & \multicolumn{1}{c|}{40.926}           & 39.722           \\ \midrule
\multicolumn{1}{|l|}{\textbf{China}}   & \multicolumn{1}{c|}{-16.228}          & \multicolumn{1}{c|}{-14.228}          & -10.8            & \multicolumn{1}{c|}{-13.968}          & \multicolumn{1}{c|}{-11.403}          & -7.9757          \\ \midrule
\multicolumn{1}{|l|}{\textbf{Japan}}   & \multicolumn{1}{c|}{29.272}           & \multicolumn{1}{c|}{32.206}           & 27.869           & \multicolumn{1}{c|}{31.532}           & \multicolumn{1}{c|}{35.031}           & 30.693           \\ \midrule
\multicolumn{1}{|l|}{\textbf{World}}   & \multicolumn{1}{c|}{24.683}           & \multicolumn{1}{c|}{26.683}           & 24.266           & \multicolumn{1}{c|}{26.943}           & \multicolumn{1}{c|}{29.508}           & 27.091           \\ \bottomrule
\end{tabular}
\caption{Estimated dynamic model comparison for the log returns of environmental indices based on the AIC and BIC (Model 1: AR(1)-ARCH(1), Model 2: AR(1)-GARCH(1,1), and Model 3: AR(1)-EGARCH(1,1))}\label{AIC_BIC}
\end{table}
To forecast the dynamic log returns for each DEI in 2022 ($R_{2022}$), we use the simulated scenarios and the estimated parameters of the univariate time series model that provides the best fit for each country (AR(1)-ARCH(1), AR(1)-GARCH(1,1), or AR(1)-EGARCH(1,1)). We generate $S$ dynamic log returns for each DEI using the estimated model parameters and the Monte Carlo simulation approach described earlier.

To obtain the dynamic asset prices for each country in 2022, we simulate $S$ dynamic log returns for 2022 using the estimated model parameters and the same Monte Carlo simulation approach. We use these simulated dynamic log returns as the basis for computing the dynamic asset prices for each country.

In summary, we use econometric models to simulate dynamic log returns for each country for the year 2022, based on their historical log returns from 2009 to 2021. The joint dependence of the log returns of all indices is modeled using a multivariate NIG distribution for the sample innovations. Our approach involves the dynamic econometric modeling of the indices as time series, which enables us to provide an asset valuation risk analysis.

It is worth noting that historical time series analyses alone are not sufficient for dynamic asset pricing, especially in option pricing. However, our econometric models are designed to be consistent with dynamic asset pricing theory, which allows us to value the indices as financial assets and price financial contracts, such as insurance instruments for the environmental indices.

\section{Regression and Jensen's Alphas of the Environmental Indices With Respect to the Global Financial Environmental Index} \label{sec:RR}


\
\
\
As we mentioned previously, the primary goal of the paper is to create a global financial environmental index that considers various economic factors that affect the environment. The previous sections have elaborated on the process used to develop this index. In this section, the focus is on analyzing the relationship between the DEIs of individual countries and the global DEI. To accomplish this task, a regression analysis is used to identify the linear relationships between the individual DEIs and the global DEI. This analysis enables a better understanding of how the financial environment of a country interacts with the global financial environment and provides insight into the complex relationships between economic factors and the environment, which can inform policy decisions and help create a more sustainable future. Policymakers can use this information to make informed decisions about economic policies that impact the environment.

In order to identify the pairwise linear dependencies between each DEI and the global DEI, a regression analysis is utilized in the following manner:
\begin{equation} \label{Eq:RR}
	r_g = \alpha_g + \beta_g \, r + e_g,
\end{equation}
where $r_g$ and $r$ represent the log returns of the exponentially transformed DEI of country $g$ and the global index, respectively. The parameters $\alpha_g$, $\beta_g$, and $e_g$ refer to the intercept, gradient, and random error of the regression line, respectively.

The ordinary least squares (OLS) method is commonly used to estimate the parameters $\alpha$ and $\beta$ when the errors follow a normal distribution. However, this assumption may not be valid if there are outliers or influential observations with high leverage. To address this issue, the robust regression (RR) method has been developed. This method assigns optimal weights to each data point through iteratively reweighted least squares, thus reducing the impact of unusual observations \citep{knez1997robustness, hu2019modelling}. OLS is preferred for short-term predictions, as it excludes outliers, while RR is more suitable for long-term forecasts in general. By using the RR method, we can obtain more reliable estimates of the parameters $\alpha$ and $\beta$, even when there are outliers and influential observations.

To estimate the parameters for the log returns of each DEI and the global DEI, we use the robust regression (RR) method. We perform this analysis using data generated for the year 2022 through time series modeling, as explained in Section \ref{sec:TS}, using dynamic regression techniques. By employing the RR method, we can gain a more comprehensive understanding of the relationship between the individual DEIs and the global DEI. This can provide valuable insights for policymakers and other stakeholders, allowing them to make informed decisions.

Table \ref{reg_world} shows the results of a regression analysis examining the relationship between the environmental finance indices (DEIs) of individual countries and the global DEI. The global DEI is the response variable, and the DEI of each country is the independent variable.

The intercept coefficient ($\alpha_g$) represents the global DEI when the DEI of each country is zero. In this case, the intercept is not significant for most countries, indicating that the global DEI is not significantly different from zero when the DEIs of individual countries are zero. The adjusted R-squared value for each country's regression model indicates the proportion of the global DEI variability that is explained by the DEI of each country. The adjusted R-squared values range between -0.026 and 0.886, suggesting that the DEIs of some countries have a stronger relationship with the global DEI than others. A negative adjusted R-squared value implies that the model is a poor fit for the data and that the independent variables are not useful for predicting the dependent variable.

The gradient coefficient ($\beta_g$) represents the change in the global DEI caused by a one-unit increase in the DEI of each country. A positive gradient coefficient indicates that an increase in the DEI of a country is associated with an increase in the global DEI, while a negative gradient coefficient indicates the opposite. The gradient coefficients are significant for most countries, indicating that the DEIs of individual countries are significant predictors of the global DEI. Specifically, Canada, France, Germany, the UK, and the US have a positive relationship between their DEIs and the global DEI, indicating that an increase in their DEIs is associated with an increase in the global DEI. However, Australia has a negative relationship with the global DEI, suggesting that an increase in its DEI is associated with a decrease in the global DEI. Overall, these results suggest that the DEIs of individual countries are important predictors of the global DEI, and the relationship between the two varies across countries.

\begin{table}[]
\centering
\begin{tabular}{@{}lccccc@{}}
\toprule
 Country         & $\alpha_g$  & $p${-}value &  $\beta_g$  & $p${-}value   & R-squared \\ \midrule
Australia & 0.273  & 0.049     & -0.708 & 0.012$^{*}$  & 0.369     \\
Brazil    & 0.120  & 0.443     & 0.228  & 0.428     & -0.026    \\
Canada    & 0.142  & 0.047     & 0.476  & 0.000$^{***}$ & 0.809     \\
China     & -0.466 & 0.087     & 1.617  & 0.017$^{*}$   & 0.340     \\
France    & 0.111  & 0.158     & 0.545  & 0.000$^{***}$ & 0.752     \\
Germany   & 0.027  & 0.792     & 0.715  & 0.001$^{***}$ & 0.558     \\
India     & -0.189 & 0.309     & 0.974  & 0.024$^{*}$  & 0.304     \\
Japan     & 0.144  & 0.242     & 0.480  & 0.013$^{*}$  & 0.364     \\
UK        & 0.154  & 0.009$^{***}$ & 0.475  & 0.000$^{***}$ & 0.886     \\
US        & 0.018  & 0.767     & 1.009  & 0.000$^{***}$  & 0.847     \\
\multicolumn{6}{l}{  *$p \leq 0.05$,    **$p \leq 0.01$,  ***$p \leq0.001$}     \\ \bottomrule
\end{tabular}
\caption{Results of the regression analysis examining the relationship between the DEIs of individual countries and the global DEI }\label{reg_world}
\end{table}

Here, we analyze the relationship between a country's financial environment and the performance of its stock market, specifically investigating the impact of each country's DEI on its primary stock market index. By analyzing this relationship, we can identify factors contributing to stock market performance and provide insights for policymakers. If a positive effect of the DEI on stock market performance is found, policymakers can use this information to develop policies that promote a healthier financial environment. Ultimately, this could lead to a stronger stock market performance and a more robust economy. 

This analysis enables us to better understand how the financial environment of a country interacts with its stock market and provides insights into the complex relationships between economic factors and the environment. Ultimately, this information can inform policy decisions and help create a more sustainable future. Policymakers can use this knowledge to make informed decisions about economic policies that positively impact the environment.

To achieve our research objective, we utilize a regression analysis to identify the linear relationships between the individual DEIs and the stock market of each country. We consider 10 major countries in our analysis: Australia (S\&P/ASX 200), Brazil (Bovespa Index), Canada (S\&P/TSX Composite Index), China (Shanghai Composite Index), France (CAC 40 Index), Germany (DAX Performance Index), India (BSE Sensex), Japan (Nikkei 225 Index), the UK (FTSE 100 Index), and the US (S\&P 500 Index).

Table \ref{Reg_stock_in} displays the regression results for the stock market performance and DEI, where $\alpha$ and $\beta$ represent the estimated intercept and gradient coefficients, respectively. The table shows the $p$-values and adjusted R-squared values for each country. The analysis provides a useful framework for understanding the complex relationship between the financial environment and stock market performance and can inform policy decisions that impact them both. 

The results show that the DEI has a significant positive effect on the stock market performances of China ($\beta$ = 0.091, $p$ = 0.65), India ($\beta$ = -0.033, $p$ = 0.92), Germany ($\beta$ = 1.906, $p$ = 0.002), Japan ($\beta$ = 2.227, $p$ = 0.002), and the US ($\beta$ = 1.676, $p$ = 0.001). These findings suggest that a healthy financial environment is crucial for stock market performance. However, the impact of the DEI on the stock market performances of Australia, Brazil, Canada, France, and the UK is not significant. Therefore, policymakers in these countries may need to consider other factors that contribute to stock market performance.

\begin{table}[]
\centering
\begin{tabular}{@{}llllll@{}}
\toprule
 Country          & $\alpha$  & $p${-}value & $\beta$  & $p${-}value & R-squared \\ \midrule
Australia & 0.194  & 0.144      & -1.193 & 0.095      & 0.150     \\
Brazil    & 0.046  & 0.750      & 0.897  & 0.103      & 0.140     \\
Canada    & -0.051 & 0.848      & 2.379  & 0.100      & 0.143     \\
China     & 0.376  & 0.000$^{***}$ & 0.091  & 0.650      & 0.064    \\
France    & 0.059  & 0.806      & 1.253  & 0.300      & 0.013     \\
Germany   & 0.066  & 0.570      & 1.906  & 0.002$^{**}$ & 0.521     \\
India     & 0.345  & 0.004$^{**}$ & -0.033 & 0.920      & 0.082    \\
Japan     & -0.079 & 0.569      & 2.227  & 0.002$^{**}$ & 0.538     \\
UK        & -0.094 & 0.774      & -1.655 & 0.526      & 0.046    \\
US        & 0.006  & 0.952      & 1.676  & 0.001$^{***}$& 0.550     \\
\multicolumn{6}{l}{  *$p \leq 0.05$,    **$p \leq 0.01$,  ***$p \leq0.001$}     \\ \bottomrule
\end{tabular}
\caption{Results of the regression analysis examining the relationship between the DEIs of individual countries and their stock markets}\label{Reg_stock_in}	
\end{table}

According to Table \ref{Reg_stock_in}, changes in the environmental indices have a significant impact on the stock markets of the United States, Japan, and Germany. The regression models exhibit high adjusted R-squared values for these countries, indicating a strong correlation between the changes in their environmental indices and stock market performance. Moreover, the testing of beta in the regression models reveals that the environmental indices of these countries have a positive impact on their respective stock markets. Specifically, positive values of beta suggest that an increase in the environmental index corresponds to an increase in stock market performance. 

However, for the other countries included in this study, the R-squared values in the regression models are relatively small. Therefore, we cannot conclude that changes in their environmental indices have a significant impact on the performance of their respective stock markets. These findings suggest that policymakers in these countries may need to consider other factors beyond the environmental indices that contribute to stock market performance.


\subsection{Performance evaluation metrics}
\
\
\
Evaluation metrics are widely used in finance to evaluate the performance of financial instruments or investment strategies. These metrics enable investors and analysts to compare different investments or strategies and allocate their resources accordingly. Commonly used evaluation metrics include return and risk measures (such as the standard deviation, beta, value at risk (VaR), and conditional value at risk (CVaR)), the Sharpe ratio, the Sortino ratio (STAR ratio), Jensen's alpha, the drawdown, the information ratio, the tracking error, and the Rachev ratio \citep{cheridito2013reward}.

To evaluate the risk-return tradeoff of different DEI scores among countries, we use a selected set of metrics, including Jensen's alpha, the Sharpe ratio, the Sortino ratio, and the Rachev ratio. These ratios can be used to compare the risk-adjusted performances of different financial instruments based on environmental indices, such as insurance products or ETFs, in different countries. A higher Jenson's alpha value, Sharpe ratio, Sortino ratio, or Rachev ratio indicates a better risk-adjusted performance of the instrument or strategy. By analyzing these ratios, investors and analysts can make informed decisions about which financial instruments provide the best risk-return tradeoff.

It is important to consider multiple evaluation metrics and other factors such as liquidity, marketability, and management costs when making investment decisions. This approach can help provide a more comprehensive assessment of the performance and risk of financial instruments or investment strategies. By using a combination of evaluation metrics and other relevant factors, investors and analysts can make better-informed decisions in the pursuit of their investment goals.

Jensen's alpha is a measure of the average return on an investment or portfolio compared to the portfolio benchmark \citep{jensen1968performance}. In a market equilibrium, Jensen's alpha is zero. However, according to \cite{soros2015alchemy}, the real financial market is always in a pre-equilibrium state, which leads to a nonzero Jensen's alpha. This means that market prices tend to fluctuate, which can result in positive or negative Jensen's alpha values. In our analysis, we use Jensen's alpha to assess the maximum possible return on each country's DEI and the performance of each country's DEI relative to the global DEI. By doing this, we can identify potential opportunities to optimize portfolio returns and gain insights into the factors that contribute to market performance.

Table \ref{Jensen_alpha} provides Jensen's alpha values for the DEI of each country, which can be used to evaluate the risk-adjusted returns of these indices. The alpha values indicate the excess return of each country's DEI compared to the global DEI after adjusting for market risk. The results show that there are significant differences in the risk-adjusted returns of DEIs across countries.

For instance, the alpha values for China, India, Japan, Germany, and Canada are all positive and relatively large; these countries have values of $2.646$, $1.616$, $1.181$, $1.095$, and $0.769$, respectively. This suggests that the risk-adjusted returns of the DEI for these countries are significantly higher than that of the global DEI. On the other hand, Brazil and the UK have negative alpha values of $-0.636$ and $-0.309$, respectively, indicating that the global DEI outperforms these countries in terms of risk-adjusted returns. The results also show that the returns of France and the US are roughly similar to that of the global DEI; these countries have alpha values of $-0.007$ and $-0.191$, respectively.

Overall, the results presented in Table \ref{Jensen_alpha} provide valuable insights into the risk-adjusted returns of each country's DEI and can inform investment decisions aimed at optimizing portfolio returns. However, it is important to note that investors should consider other factors such as market volatility, liquidity, and diversification when making investment decisions.

\begin{table}[h]
\centering
\begin{tabular}{@{}lcc@{}}
\toprule
Country & Alpha   \\ \midrule
Brazil  & -0.636 \\
UK      & -0.309 \\
US      & -0.191 \\
France  & -0.007 \\
Canada  & 0.769  \\
Germany & 1.095   \\
Japan   & 1.181  \\
India   & 1.616  \\
China   & 2.646  \\ \bottomrule
\end{tabular}
\caption{Jensen's alpha (CAPM) for each country's index}
\label{Jensen_alpha}
\end{table}

The Sharpe ratio is a widely used evaluation metric in finance that can be applied to assess the risk-adjusted performance of financial instruments or investment strategies. To evaluate the risk-adjusted performance of financial instruments or investment strategies based on their performance relative to DEIs in different countries, we can use the Sharpe ratio. This ratio assesses the excess return of a DEI over the risk-free rate, taking into account the level of risk taken to achieve that return. By comparing the Sharpe ratios of DEIs in different countries, we can identify which DEIs offer the best risk-adjusted performance. This information can be used to inform investment decisions and guide efforts toward sustainable development.

The Sharpe ratio is a widely used measure of the risk-adjusted performance in finance. It assesses the excess return of an investment strategy or financial instrument over the risk-free rate, taking into account the level of risk taken to achieve that return. A higher Sharpe ratio indicates a better risk-adjusted performance.

From Table \ref{Sharpe_ratio}, we can see that China has the highest Sharpe ratio of $1.8453$, followed by India, which has a Sharpe ratio of $1.0527$. These two countries have the best risk-adjusted performance in terms of the DEI. On the other hand, the UK has a negative Sharpe ratio of $-0.0195$, indicating that its DEI has underperformed compared to a risk-free investment. 

Australia, Germany, and the US have positive Sharpe ratios, indicating that their DEIs have provided a positive excess return compared to a risk-free investment, but their risk-adjusted performances are still lower than those of China and India. Brazil, France, Canada, and Japan have even lower Sharpe ratios, indicating a lower risk-adjusted performance compared to the other countries.

\begin{table}[]
\centering
\begin{tabular}{@{}lc@{}}
\toprule
 Country         & Sharpe ratio \\ \midrule
China     & 1.8453      \\
India     & 1.0527      \\
Australia & 0.3767      \\
Germany   & 0.2870      \\
US        & 0.2575      \\
Brazil    & 0.1963      \\
France    & 0.0725      \\
Canada    & 0.0041      \\
Japan     & 0.0002      \\
UK        & -0.0195     \\ \bottomrule
\end{tabular}
\caption{Sharpe ratios of the DEIs of different countries}
\label{Sharpe_ratio}
\end{table}

These results could be used to guide investment decisions for investors interested in investing in DEIs across different countries. However, it is important to note that while the Sharpe ratio is a widely used measure of the risk-adjusted performance, it has the following limitation: It treats all deviations from the mean return as risk. This means that it penalizes both positive and negative deviations equally, even though investors are typically more concerned with negative deviations. Thus, investors may want to consider using alternative measures such as the Sortino ratio, which focuses on the downside risk and provides a more accurate measure of the risk-adjusted performance for those who are more risk-averse. Additionally, investors should consider other factors such as market volatility, liquidity, and diversification when making investment decisions.

The Sortino ratio addresses this limitation by only considering the deviation of returns below a certain target return, which is typically the risk-free rate. This allows investors to focus on the downside risk and better evaluate the risk-adjusted performance of an investment strategy or financial instrument.

In other words, the Sortino ratio is a variation of the Sharpe ratio that takes into account the downside risk, which is a more relevant measure of risk for investors. By only considering the downside deviation from the target return, the Sortino ratio provides a more accurate measure of the risk-adjusted performance, especially for investors who are more risk-averse.

From Table \ref{Sortino_ratio}, we can see that China has the highest Sortino ratio of $111.53$, followed by India, which has a Sortino ratio of $2.1476$. These two countries have the best risk-adjusted performance in terms of the DEI. 

Australia, Germany, the US, and Brazil also have positive Sortino ratios, indicating that their DEIs have provided a positive excess return compared to the target return, but their risk-adjusted performances are still lower than those of China and India.

France, Canada, Japan, and the UK have even lower Sortino ratios, indicating a lower risk-adjusted performance compared to the other countries. The negative Sortino ratio of the UK indicates that its DEI has underperformed compared to the target return.

These results could be used to guide investment decisions for investors interested in investing in DEIs across different countries, particularly for those who are more risk-averse and focus on the downside risk. However, it is important to note that the Sortino ratio, like the Sharpe ratio, is not the only measure of the risk-adjusted performance, and investors should consider other factors such as market volatility, liquidity, and diversification when making investment decisions. Therefore, in addition to the Sortino ratio and Sharpe ratio, investors may also want to compute the Rachev ratio, which is a measure of the downside risk of an investment and can be useful for investors who are concerned about extreme negative events in the market.

\begin{table}[]
\centering
\begin{tabular}{@{}lc@{}}
\toprule
 Country         & Sortino ratio \\ \midrule
China     & 111.53        \\
India     & 2.1476        \\
Australia & 0.6212        \\
Germany   & 0.4071        \\
US        & 0.3679        \\
Brazil    & 0.3346        \\
France    & 0.0861        \\
Canada    & 0.0048        \\
Japan     & 0.0003        \\
UK        & -0.0224      \\ \bottomrule
\end{tabular}
\caption{Sortino ratios of the DEIs of different countries}
\label{Sortino_ratio}
\end{table}

To evaluate and compare the risk-return tradeoff of different DEI scores among countries, we can use the Rachev ratio, or R-ratio, as a measure of performance. The R-ratio is a risk-return measure that allows us to assess the upside potential relative to the downside risk for each country's DEI score.
It is a valuable tool for evaluating the performance of an investment portfolio, and it can be adapted to be used with DEI scores. By calculating the R-ratio for each country's DEI score, we can compare their expected returns relative to the downside risk\footnote{The downside risk is the chance that an investment may incur losses or underperform in comparison to its expected return, especially during unfavorable market conditions. It stands in contrast to the upside potential, which describes the potential for an investment to generate higher-than-expected returns. In the realm of finance, the downside risk is commonly assessed using statistical measures such as the value at risk (VaR), conditional value at risk (CVaR), expected tail loss (ETL), or other methods that account for negative returns. These metrics help investors understand the potential risks of their investments and make informed decisions.}.
This enables us to assess which country's DEI score has a higher expected return relative to its downside risk compared with the DEI scores of other countries. 

The R-ratio is designed to capture the right tail reward potential compared to the left tail risk at a rarity frequency $q$, which is a quantile level defined by the user. The ratio is defined as the expected tail return (ETR) in the best $q\%$ of cases divided by the expected tail loss (ETL) in the worst $q\%$ of cases.


In its original version, which was introduced in 2004, the Rachev ratio is defined as follows:
\begin{equation}
\label{R_ratio}
  \rho \left({x'r}\right)={\frac {CVa{R_{(1-\alpha )}}\left({{r_{f}}-x'r}\right)}{CVa{R_{(1-\beta )}}\left({x'r-{r_{f}}}\right)}}  ={\frac {ET{L_{\alpha }}\left({{r_{f}}-x'r}\right)}{ET{L_{\beta }}\left({x'r-{r_{f}}}\right)}},
\end{equation}
where 
$\alpha$  and 
$\beta$  belong to
$\left({0,1}\right)$, and in the symmetric case, 
$\alpha =\beta $. $r_f$ is the risk-free rate of return and 
$ x'r$ represents the portfolio return.
A higher Rachev ratio indicates a portfolio with a higher expected return relative to its downside risk, which is generally considered more desirable.



Table \ref{Rachev_Ratio} presents the R-ratio values for each country's DEI score in the symmetric case, where $\alpha=\beta=0.5$. By comparing the R-ratios of the DEIs, we can identify which countries offer a better risk-return tradeoff. A higher Rachev ratio indicates a lower level of downside risk, while a lower Rachev ratio indicates a higher level of downside risk.

In the table, we can see that the Rachev ratios for the UK, Japan, and Canada are greater than $0.9$, indicating that these countries have relatively low levels of downside risk or a better risk-return tradeoff compared to the other countries. This means that these countries have a higher expected return relative to their downside risk compared to the other countries.

The Rachev ratios for France, Brazil, the US, Germany, and Australia are between $0.5$ and $0.9$, indicating moderate levels of downside risk. Finally, the Rachev ratios for India and China are negative\footnote{A negative Rachev ratio is an unusual result that indicates a highly skewed distribution with a long left tail. In other words, the distribution of returns is heavily skewed to the left, indicating that there is a high probability of large negative returns. This indicates that the downside risk is much greater than the upside potential for these countries, which is a cause for concern for investors.}, indicating that the downside risk is much higher than the upside potential. This means that these countries have a lower potential for extreme positive returns compared to the risk of extreme losses (negative returns). This indicates a less favorable risk-return tradeoff for these countries' DEI scores.

\begin{table}[h]
\centering
\begin{tabular}{@{}lc@{}}
\toprule
Country          & Rachev ratio \\ \midrule
UK        & 1.0587       \\
Japan     & 0.9994       \\
Canada    & 0.9889       \\
France    & 0.7723       \\
Brazil    & 0.5862       \\
US        & 0.4637       \\
Germany   & 0.4048       \\
Australia & 0.3404       \\
India     & -0.2435      \\
China     & -0.5550      \\ 
\bottomrule
\end{tabular}
\caption{Rachev ratio for each country index for $\alpha =\beta = 0.5$}\label{Rachev_Ratio}
\end{table}

\newpage
\section{Efficient Frontier of the Markets of a Country's Environmental Risk Measures
} \label{sec:EF}



\
\
\
\
Modern portfolio theory, introduced by Markowitz in 1952, aims to identify the optimal set of daily weights $w$ for a portfolio consisting of $n$ risky assets. The objective of this optimization is to minimize the risk of the portfolio's return for a specific day while achieving a desired level of the expected return, denoted as $r_p$. The level of the expected return selected by an investor reflects their appetite for risk, with higher values indicating a greater willingness to accept risk.

Two commonly used optimization techniques are mean-variance and mean-CVaR optimization. The former method minimizes the portfolio variance, $\sigma_p$, while the latter minimizes the portfolio CVaR, which is denoted by CVaR$_{p,\alpha}$; both methods utilize the variance and CVaR as risk measures.

The portfolio can be represented by the daily return values $r(t) = (r_1(t), r_2(t), ..., r_n(t))$, with a portfolio mean and standard deviation of $\bar{r} = (\bar{r_1}, \bar{r_2}, ..., \bar{r_3})$ and $\sigma_p=(\sigma_1,\sigma_2, ..., \sigma_n)$, respectively.

The optimization problem can be formulated as follows:

\begin{equation}
\text{minimize}\,\, w' \sigma_p w \,\, \text{subject to} \,\, \bar{r}w = r_p \,\,\text{and} \,\,\sum_{i=1}^{n} w_i = 1,
\end{equation}
where $w' \sigma_p w$ represents the portfolio variance, $\bar{r}w = r_p$ denotes the desired level of the expected return, and $\sum_{i=1}^{n} w_i = 1$ ensures that the weights add up to one.

Another important aspect of portfolio optimization is the efficient frontier (EF), which represents the best solutions for different levels of the expected return, denoted as $r_p$, and portfolio variance, denoted as $\sigma_p$. The EF is the portion of the portfolio frontier where the projected mean returns exceed the desired level of the expected return.

Consider a portfolio consisting of $n$ risky assets with daily return values $r_p$, with an expected risk-adjusted mean return of $E(r_p)$ and a risk measure of $V(r_p)$. The portfolio optimization problem can be formulated as follows:

\begin{equation}
\text{min}w (-\gamma E(r_p) - (1-\gamma) V(r_p)) \,\, \text{subject to} \,\,\sum{i=1}^{n} w_i = 1,
\end{equation}

where $\gamma \in [0,1]$ represents the risk-aversion parameter that determines the position along the EF. Specifically, $\gamma = 0$ corresponds to the minimum-risk portfolio.

By optimizing the portfolio using this equation, we can identify the optimal set of daily weights that minimize the risk of the portfolio's return while achieving the desired level of the expected return. This helps investors to make informed decisions about their investments and balance the risk and return tradeoff to meet their investment objectives.

For the environmental financial index of each country, two risk measures are employed: the standard deviation and CVaR \citep{uryasev2001conditional}. The optimization process involves three risk measures: the variance, CVaR$_{p,0.05}$, and CVaR$_{p,0.01}$. While the variance is used as a central risk measure, CVaR$_{p,0.05}$ and CVaR$_{p,0.01}$ are utilized as tail risk measures. It is worth noting that the variance is not a coherent risk measure, whereas CVaR$_{p,0.05}$ and CVaR$_{p,0.01}$ are coherent risk measures  \citep{artzner1999coherent}.

\begin{figure}[h!]
	\centering
	\includegraphics[width=0.75\textwidth]{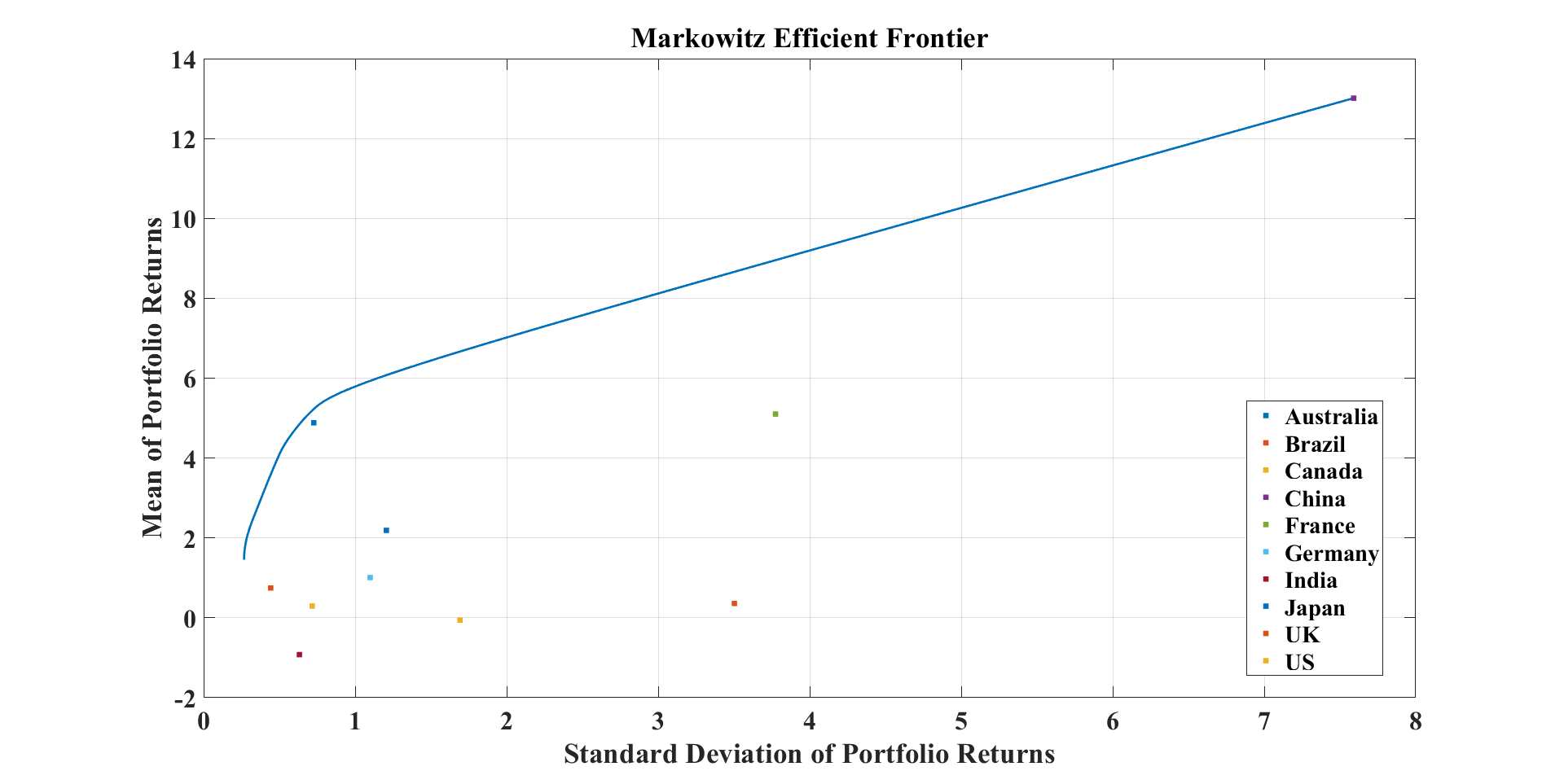} 
	\caption{Markovitz efficient frontier}
	\label{Fig_Mark}			
\end{figure}

Figure \ref{Fig_Mark} displays the EFs computed for the global DEI. To generate efficient borders for the US dynamic and historical DEI index, we utilized the mean-variance risk measure. Each EF was plotted using the set of equally spaced values $\gamma = \{0, 0.01, ..., 0.99\}$. Notably, the standard deviation sharply increased in the EF starting from 0.8. Among the countries, India's and Canada's indices have negative returns, while China's index has the highest return and the highest risk. Following China, the highest returns are from France and Japan. However, when comparing the variance of these countries, Japan proves to be a favorable choice for investing, as its variance is low. On the other hand, Brazil is among the countries with the lowest returns, while its variance is very high, making it the worst option for investing.

\begin{figure}[h!] 	
	\centering
	\subfigure[]{\includegraphics[width=0.45\textwidth]{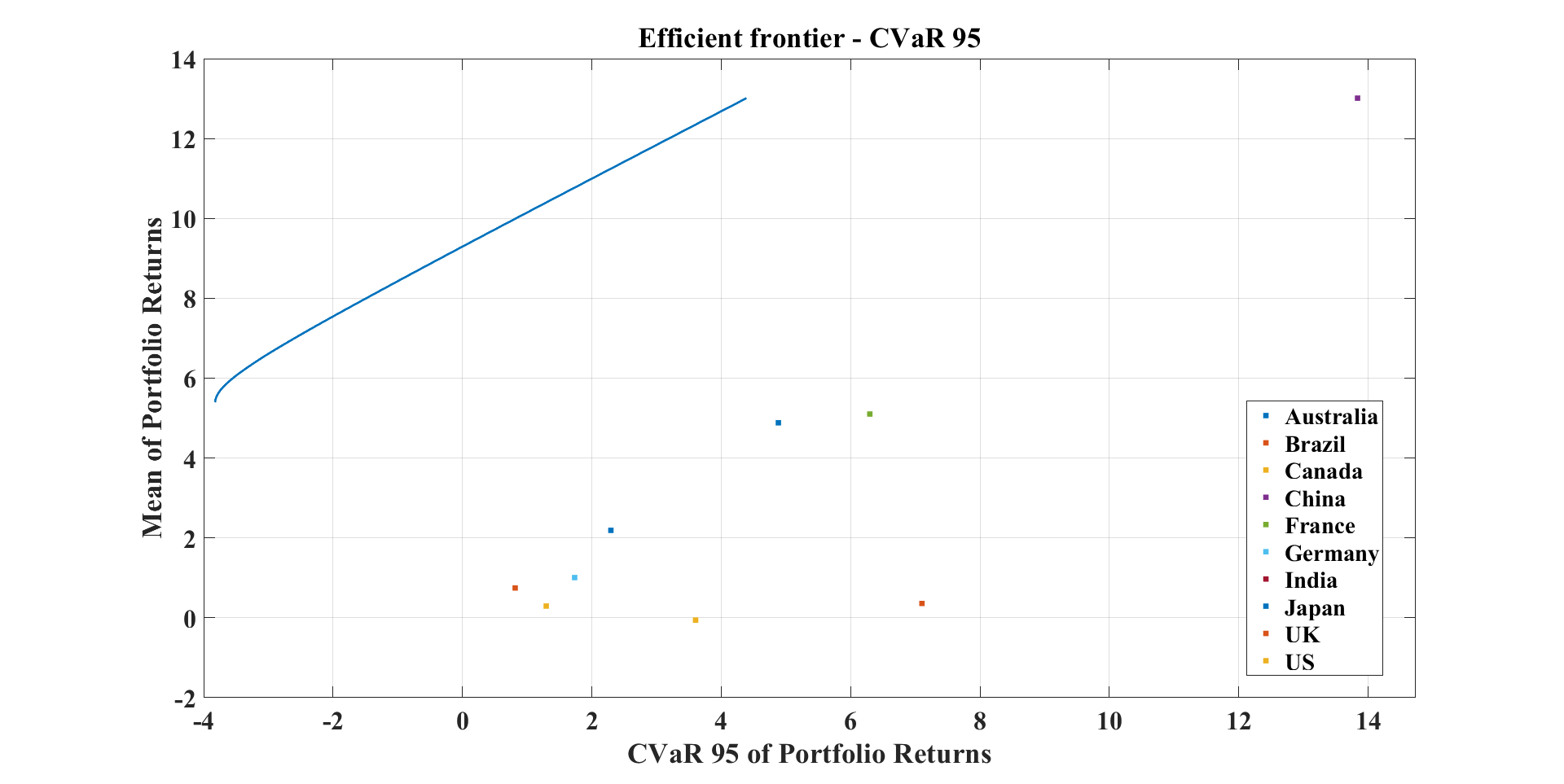}}
	\subfigure[]{\includegraphics[width=0.45\textwidth]{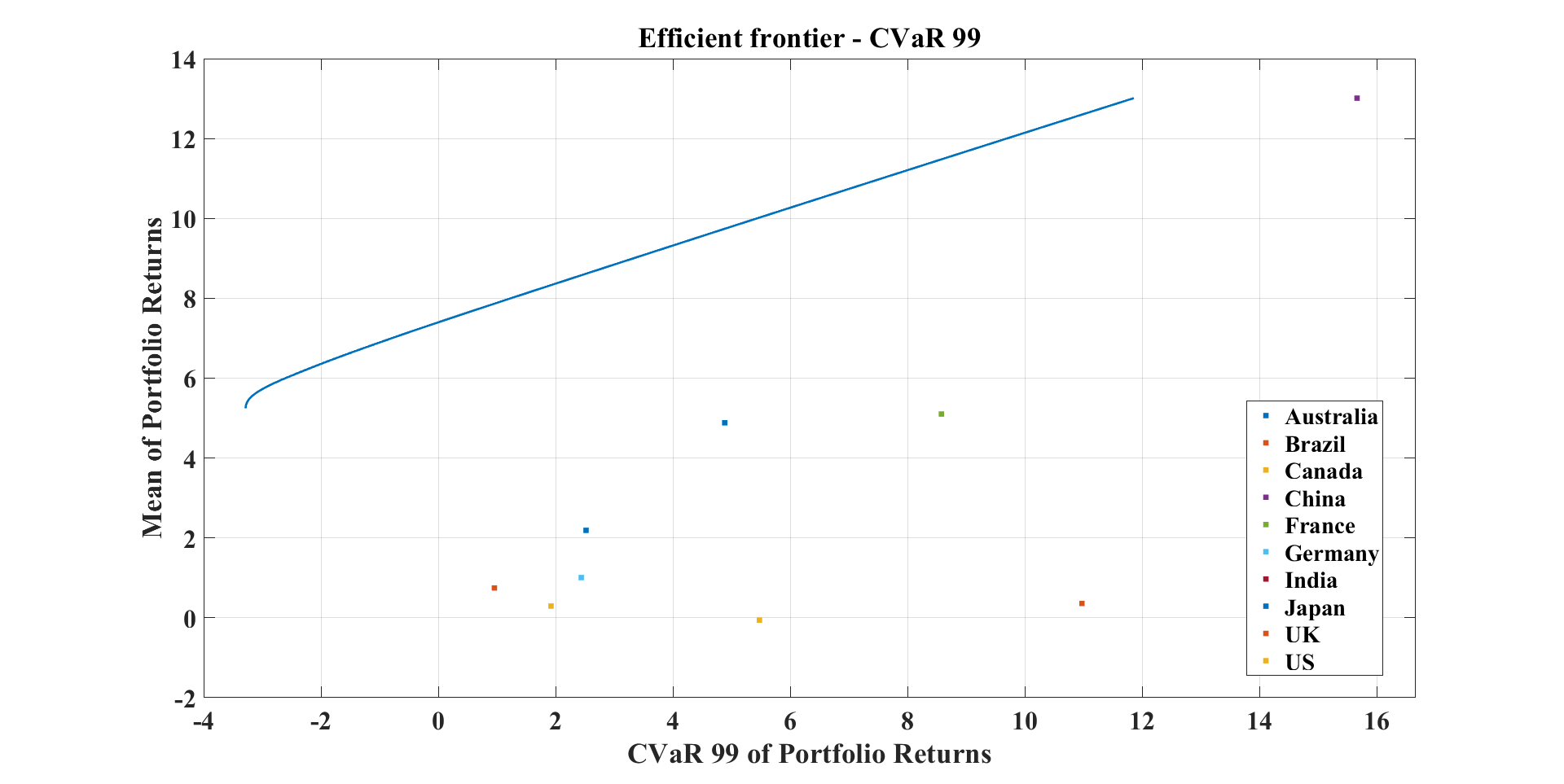}}
	\caption{Conditional value-at-risk portfolio optimization: (a) CVaR$_{p,0.05}$ and (b) CVaR$_{p,0.01}$ EFs}	
	\label{Fig_CVaR}			
\end{figure}

Figures \ref{Fig_CVaR}(a) and \ref{Fig_CVaR}(b) depict a reproduction of Figure \ref{Fig_Mark} that utilizes the tail risk measures CVaR$_{p,0.05}$ and CVaR$_{p,0.01}$. It is worth noting that CVaR$_{p,0.01}$ changes more smoothly than CVaR$_{p,0.05}$, and the variation in the behavior of the CVaR$_{p,0.05}$ EF is more pronounced than that of the CVaR$_{p,0.01}$ EF.

When comparing the behaviors of the CVaR EFs with the mean-variance EF, it is clear that they exhibit overall qualitative similarities. However, the CVaR risk measures are more volatile.

\newpage

\section{Pricing the Options on Environmental Indices} \label{sec:OP}
\
\
\
\
This section introduces a financial model for pricing options on the environmental indices of individual countries and the world. To do this, 
we utilize dynamic forecasting models to evaluate future trends in environmental indices on country-by-country and global scales. As financial options are designed for hedging, risk assessment, and speculative purposes, we provide DEI option prices to financial institutions to add an additional socioeconomic dimension to their risk-return-adjusted portfolios.


Examples of traditional techniques for calculating the prices of options include the Black-Scholes-Merton model, the binomial option pricing model, the trinomial tree, the Monte Carlo simulation, and the finite difference model \citep{hull2003options, duffie2010dynamic}. However, due to the presence of heteroskedasticity and the distributional heavy-tailedness of the DEI's returns, we cannot employ the Black-Scholes-Merton model to price DEI options.

Instead, a discrete stochastic volatility-based model was developed to calculate option prices and explain some well-known mispricing events \citep{Black_1973, madan1989multinomial, carr2004time, bell2006option, klingler2013option, shirvani2020optionHu, Shirvanijod_2021_1_138}. In particular, \cite{Duan:1995} proposes the application of the discrete-time GARCH model to price options. We employ the discrete-time GARCH model with NIG innovations to calculate the fair prices of DEI options while accounting for the accuracy of the pricing performance using a volatility-based approach \citep{Duan:1995, barone2008garch, chorro2012option}.

In the standard GARCH(1,1) model, the distribution of $R_t$ for a given $F_{t-1}$ is defined on the real-world probability space ($\mathbb{P}$) as $R_t \sim NIG\left(\lambda,\frac{\alpha}{\sqrt{a_t}}, \;\frac{\beta}{\sqrt{a_t}}, \;\delta\sqrt{a_t}, \;r'_t+m_t+\mu \sqrt{a_t} \right)$, where $m_t=\lambda_0 \sqrt{a_t}-\frac{1}{2}a_t$ \citep{Blaesild_1981}. However, due to the presence of heteroskedasticity and heavy-tailedness in the DEI's returns, we cannot use the traditional approach of determining an equivalent martingale measure to obtain a constant option price \citep{Gerber:1994}. Instead, \cite{chorro2012option} discovered that using the Esscher transformation to transform the probability density of $R_t$ into the risk-neutral probability density results in $R_t$ for a given $F_{t-1}$ being distributed on the risk-neutral probability ($\mathbb{Q}$) as $R_t \sim NIG\left(\lambda,\frac{\alpha}{\sqrt{a_t}},\frac{\beta}{\sqrt{a_t}}+\theta_t,\delta \sqrt{a_t},r'_t+m_t+\mu \sqrt{a_t} \right)$, where $\theta_t$ is the solution to $MGF\left(1+\theta_t \right)= MGF\left(\theta_t \right) \, e^{r'_t}$ and $MGF$ is the conditional moment-generating function of $R_{t+1}$ given $F_{t}$.

To price the DEI call and put options, we employ Monte Carlo simulations \citep{chorro2012option} and construct future values of the DEI as follows:

1. We fit a GARCH(1,1) model with NIG innovations to $R_t$ and forecast $a_1^2$ by setting $t=1$.

2. We repeat the following substeps for $t=3,4,...,T$, where $T$ is the time to maturity of the DEI call option starting from $t=2$:
    (a) We estimate the model parameter $\theta_t$ using $MGF\left(1+\theta_t \right)= MGF\left(\theta_t \right) \, e^{r'_t}$, where $MGF$ is the conditional moment-generating function of $R_{t+1}$ given $F_{t}$ on $\mathbb{P}$.
    (b) We find an equivalent distribution function for $\epsilon_t$ on $\mathbb{Q}$.
    (c) We generate the value of $\epsilon_{t+1}$ under the assumption $\epsilon_{t} \sim NIG(\lambda,\alpha,\beta+\sqrt{a_t}\theta_t,\delta,\mu)$ on $\mathbb{Q}$.
    (d) We compute the values of $R_{t+1}$ and $a_{t+1}$ using a GARCH(1,1) model with NIG innovations.

3. We generate future values of $R_t$ for $t=1,....,T$ on $\mathbb{Q}$, where $T$ is the time to maturity. Recursively, future values of the DEI are obtained by $DEI_{t}= \exp(R_{t}) \cdot DEI_{t-1}$.

4. We repeat steps 2 and 3 10,000 ($N$) times to simulate $N$ paths to compute future values of the DEI.

For a specific strike price $K$, the approximate future value of the DEI at time $t$ is the average of the DEIs, and this price is used to determine the prices of the call ($\hat{C}$) and put ($\hat{P}$) options as follows:

\begin{equation} 
\hat{C}\left(t,T,K \right)=\frac{1}{N}\,e^{-r'_t(T-t)}\sum_{i=1}^{N} \max \left(DEI^{(i)}_T-K,0 \right), 
\end{equation}
\begin{equation} 
\hat{P}\left(t,T,K \right)=\frac{1}{N}\,e^{-r'_t(T-t)}\sum_{i=1}^{N} \max \left(K-DEI^{(i)}_T, 0 \right).
\end{equation}

The call option pricing ($\hat{C}$) helps investors plan to purchase the DEI stocks at a predetermined strike price within a predetermined time frame (time to maturity).

The calculation of the call option price ($\hat{C}$) is beneficial for investors, as it allows them to plan to purchase the DEI stocks at a predetermined strike price within a specified time frame (time to maturity). By using the Monte Carlo simulation approach and the GARCH(1,1) model with NIG innovations, we can estimate the future values of the DEI and use them to determine the option price. This information can assist investors in making informed decisions about their investment strategies.

\begin{figure}[h!] 	
	\centering
	\subfigure[]{\includegraphics[width=0.45\textwidth]{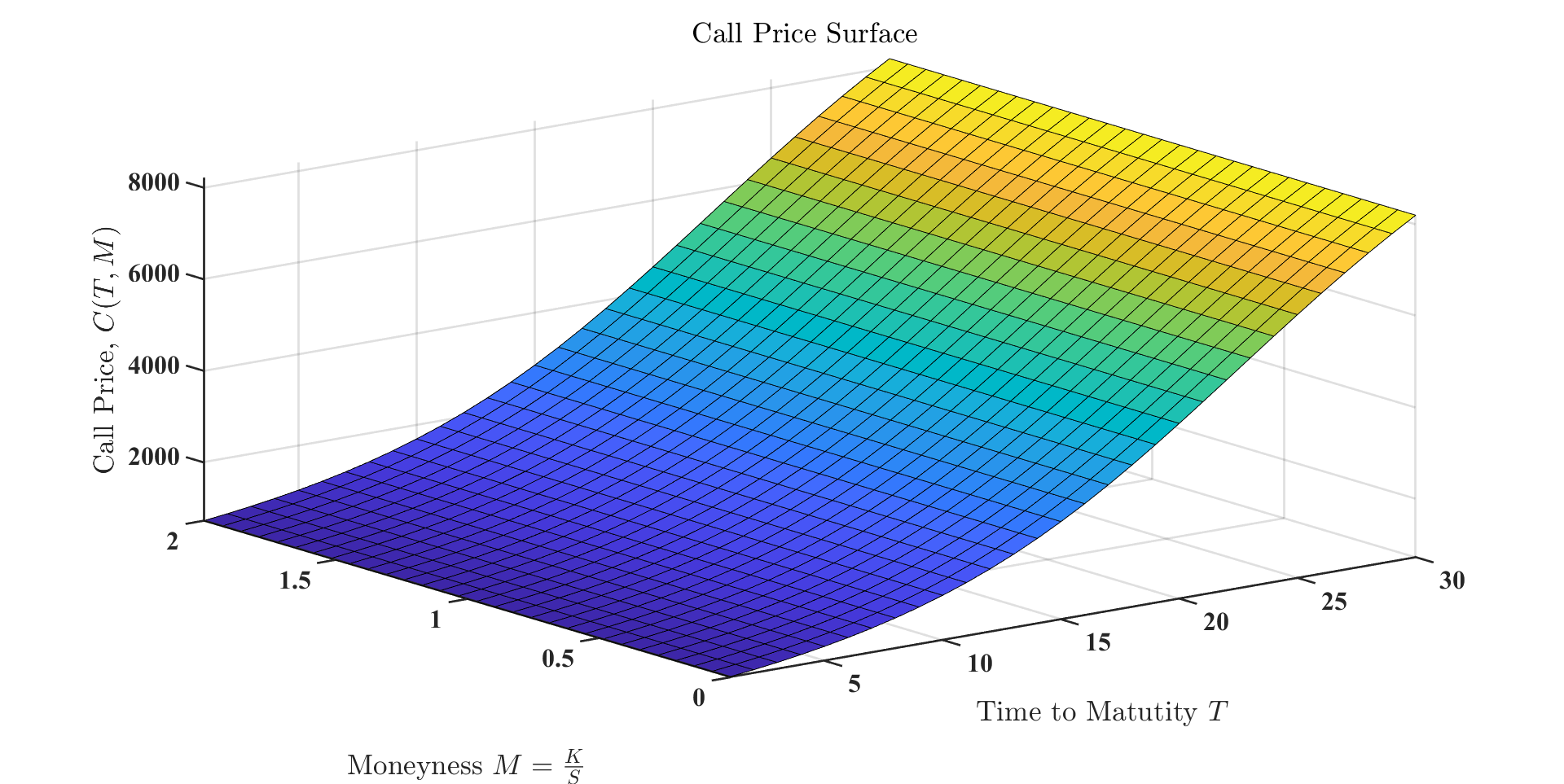}}
	\subfigure[]{\includegraphics[width=0.45\textwidth]{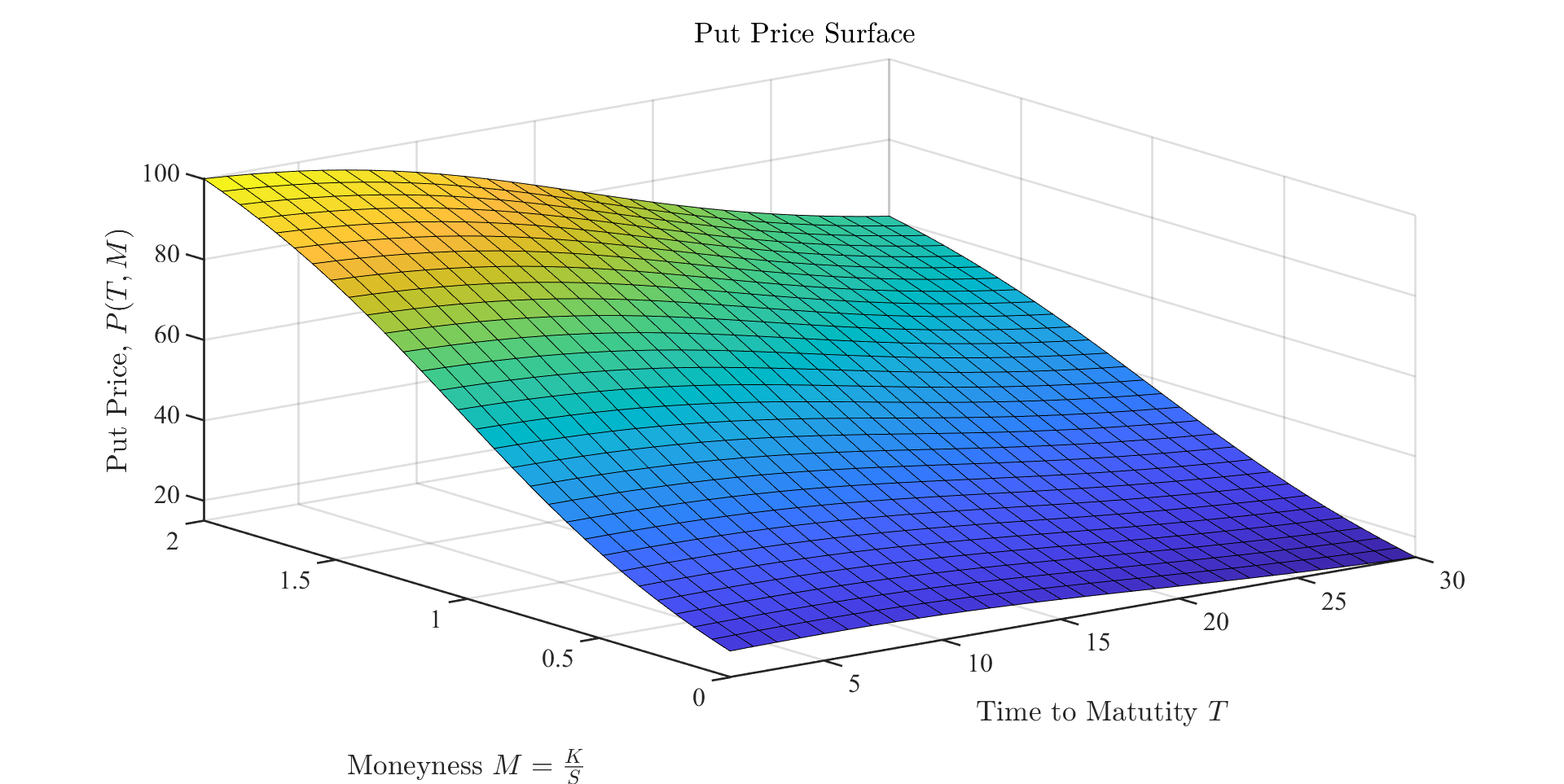}}
	\caption{Option prices for the US DEI at time $t$ for a given strike price $K$ using a $GARCH(1,1)$ model with NIG innovations: 
 (a) call prices and (b) put prices}	
	\label{Fig_option_prices}			
\end{figure}

The prices of the call and put options for the DEI are displayed in Figure \ref{Fig_option_prices}(a), and they are determined based on the moneyness $(S/K)$ and time to maturity $(T)$ of the DEI. As the strike price increases, there is a slight decrease in the price of the DEI call option, while a decrease in the strike price leads to a slight increase in the call option price. This occurs because the call option gives the holder the right to buy the underlying asset at a fixed price, and as the strike price decreases, the likelihood of the underlying asset's price exceeding the strike price and making the option profitable increases, resulting in a higher option price. Moreover, an increase in the time to maturity leads to an increase in the price of the DEI call option, indicating a potential rise in the DEI's price. These trends can provide valuable insights for investors in their decision-making process regarding the DEI options.

In Figure \ref{Fig_option_prices}(b), we provide the selling prices for the shares in our index by exploring the relationship between the put option prices ($\hat{P}$) and the strike price $(K)$, as well as the time to maturity $(T)$. It is worth noting that the put option prices are generally lower than the call option prices for the same moneyness and time to maturity values. This is because the put option gives the holder the right to sell the underlying asset (in this case, the DEI) at a fixed price, which is less valuable than the right to buy the asset at a fixed price. 

Moreover, as expected, the put option price increases as the strike price increases. This trend suggests that the put option price and the strike price may be related linearly, indicating that a higher strike price leads to a higher put option price. Additionally, an increase in the time to maturity leads to an increase in the put option price, indicating that the price of the DEI may decline in the future. These insights can assist investors in making informed decisions concerning their investment strategies related to the DEI put options and the selling prices of the DEI shares.

\begin{figure}[h!]
	\centering
	\includegraphics[width=0.75\textwidth]{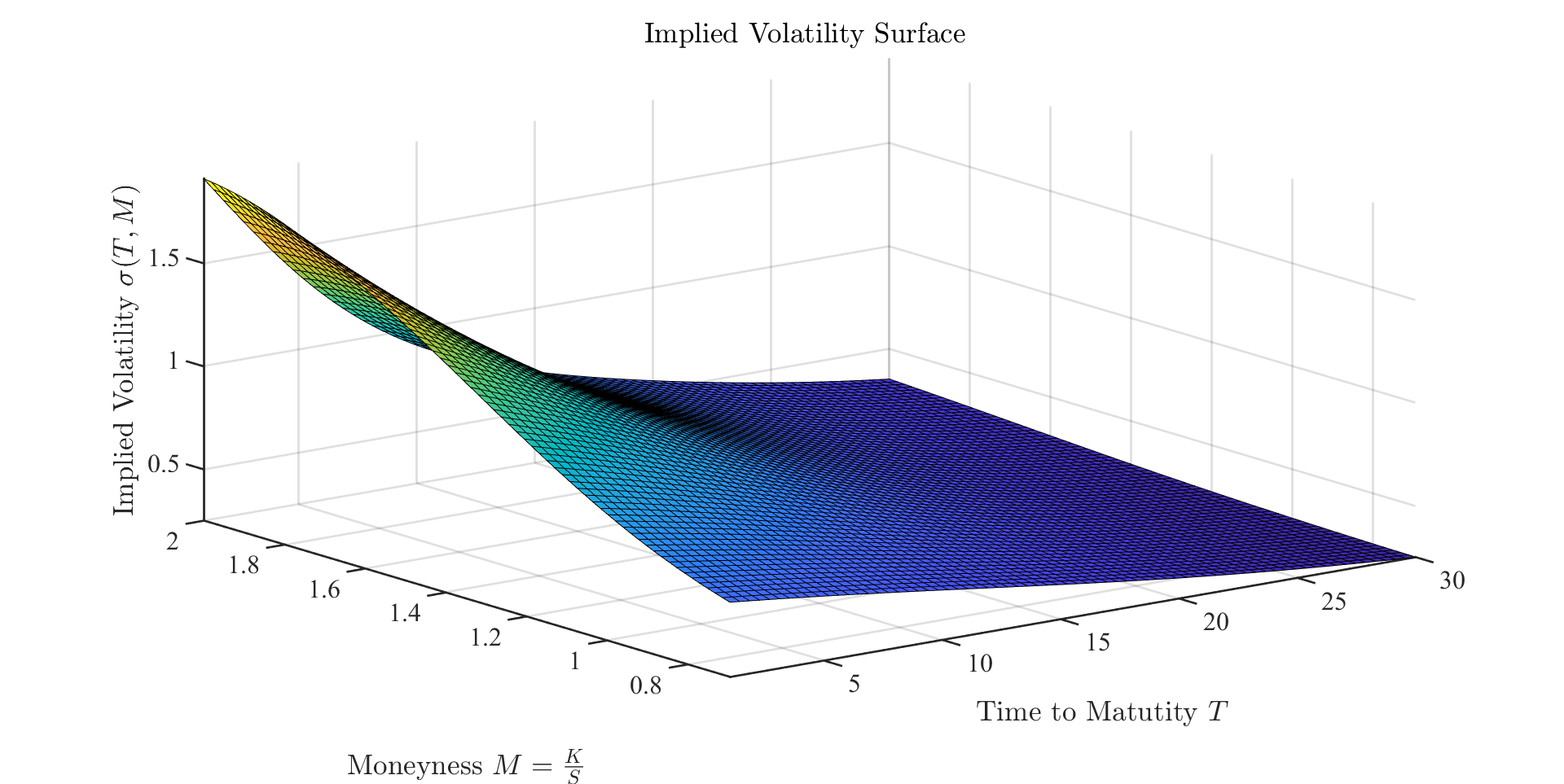} 
	\caption{US DEI implied volatilities versus the time to maturity $(T)$ and moneyness
$(M = \frac{S}{K}$, where $S$ and $K$ are the stock and strike prices, respectively) according to a $GARCH(1,1)$ model with NIG innovations}
	\label{Fig_Implied_vol}			
\end{figure}

The implied volatility is a commonly used metric in finance that is considered a reliable forecast of future volatility for the remainder of the option's life \citep{poterba1984persistence, day1988behavior, sheikh1989stock, harvey1992market}. Figure \ref{Fig_Implied_vol} displays the implied volatility for the DEI, which is calculated by using the market values of the call option contracts as a proxy for the market's expectation of an upcoming event. The implied volatility surface is constructed using the time to maturity ($T$) and moneyness ($M=S/K$, where $S$ and $K$ are the stock and strike prices, respectively) of the options. 

During periods of intense market stress, an inverted volatility grin may be observed on the implied volatility surface, with the highest implied volatilities being observed when the moneyness increases. It is also worth noting that option implied volatilities are generally higher for options with lower strike prices than for those with higher strike prices. This is because the implied volatility for upside (low-strike) equity options is often lower than the implied volatility for equity options in the money, resulting in a downward-sloping (volatility skew) graph. 

Furthermore, as the time to maturity increases, the suggested volatilities tend to converge to a constant. These option prices are primarily used for hedging purposes rather than speculation and can play a role similar to portfolio insurance \citep{portins}. The implied volatility surface can provide valuable insights for investors in their decision-making process regarding the DEI options and risk mitigation strategies.

\nc

\newpage

\section{Factor Analysis} \label{sec:FA}
\
\
\
\
Factor analysis is a statistical technique commonly used in finance and other fields to identify the underlying factors that explain the variation in a set of observed variables. In finance, factor analysis is often used to identify the common factors that influence the returns of a portfolio of assets, such as stocks, bonds, or other financial instruments. By identifying these underlying factors, investors can gain insights into the sources of risk and return in their portfolios and potentially use this information to make more informed investment decisions.

Factor analysis can provide valuable insights for policymakers and investors who are interested in promoting sustainable economic development in a global context. By using factor analysis to analyze the GDEI, we can identify which countries are leaders in sustainable economic development and which countries have room for improvement. This can inform policy decisions and investment strategies that promote sustainability and responsible economic growth.

Here, for the GDEI, we have created financial environmental indices for 10 countries based on economic factors that measure the impact of economic activities on the environment for each country. To gain a deeper understanding of the complex relationship between economic activities and the environment across these 10 countries, we conduct factor analysis on the GDEI using each country index as a factor. By doing this, we can identify the key drivers of environmental degradation and economic growth across these countries and develop policies and strategies that promote sustainable economic development.

Factor analysis can help us to identify the common factors that underlie the complex relationship between economic activities and the environment and to gain insight into which countries are most closely related to each other and which countries are most strongly associated with the overall GDEI score. By examining the factor loadings, we can gain a deeper understanding of the sources of risk and return in a portfolio of economic and environmental indicators.

When considering each country index as a factor that is used to create the GDEI, a key question that can be addressed using factor analysis is whether there are common factors that explain the observed behavior of the GDEI. Factor analysis quantifies this by measuring the relative weight or loading of each country index on the GDEI. These loadings represent the degree to which each country index influences the GDEI. In addition, factor analysis provides an estimate of the portion of the GDEI's returns that remains unexplained by the common factors, which is known as the latent error. By identifying these common indices and quantifying their influence on GDEI returns, policymakers and investors can gain valuable insights into the sources of risk and return in the GDEI, which can inform sustainable economic development strategies.

One familiar model used in finance is the single-factor CAPM of Sharpe (1970), which can be used to determine the degree to which each asset in a portfolio is influenced by the excess market return.

The CAPM equation expresses the return for asset $i$ as a function of the risk-free rate, the excess market return, and the loading of asset $i$ on the excess market return. The model assumes that there is a single factor that drives returns, namely the excess market return. The loading of asset $i$ on the excess market return is represented by $\beta_i$, and the model also includes a term for the latent error, which is the portion of the return that cannot be explained by the model. In the original CAPM, as formulated by Sharpe in the following manner, the latent error is assumed to be zero for all assets: 

\begin{equation}
r_{it} = r_{ft} + \beta_{i}r_{mt} + \epsilon_{it}, \,\,\,\,\,\, i = 1,...,N , \,\,\,\,\,\, t = 1,...,T, 
\end{equation}
where $r_{it}$ is the return for asset $i$; $r_{ft}$ is the
risk-free rate; the single factor, $r_{mt}$, is the excess (relative
to the risk-free rate) market return; $\beta_{i}$ is the loading of
asset $i$ on the excess market return; and $\epsilon_{it}$ is the
latent error for asset $i$.

The CAPM can be used to estimate the expected returns for assets and to determine whether they are overvalued or undervalued relative to their expected returns.

To construct a factor model for a portfolio such as the GDEI, one can use a vector of returns for each of the $N$ assets in the portfolio, denoted by ${\mathbf{r}_{t} = (r_{1t},\ \ ...,\ r_{Nt})}^{T}$, with a mean vector of $\mathbf{\mu}{= (\mu_{1},\ \ ...,\ \mu_{N})}^{T}$. The general factor model expresses the return of the portfolio at time $t$ as the sum of the mean return, the product of the factor loadings and a vector of the scores for each common factor at time $t$, and a vector of the latent errors at time $t$ as follows:

\begin{equation}
  \mathbf{r}_{t} = \mathbf{\mu} + \mathbf{\beta}\mathbf{f}_{\mathbf{t}} + \mathbf{\epsilon}_{t}, \,\,\,\,\,\,\,\,\,\, t = 1,...,T .  
\end{equation}

The factor scores are represented by ${\mathbf{f} = (f_{1t},\ \ ...,\ f_{mt})}^{T}$, where $m < N$, and the factor loadings are represented by the matrix $\mathbf{\beta =}\left\{ \beta_{ij} \right\}$, with $\beta_{ij}$ being the loading of asset $i$ on the $j$'th common factor. The latent errors are represented by ${\mathbf{\epsilon}_{\mathbf{t}} = (\epsilon_{1t},\ \ ...,\ \epsilon_{Nt})}^{T}$. If the latent errors are assumed to be independent, then their covariance matrix is diagonal and can be expressed as $\mathbf{D} = diag\left\{ \sigma_{1}^{2},\ ..,\ \sigma_{N}^{2} \right\}$.

The covariance of the return series $\mathbf{r}_{t}$ can be expressed as the sum of the covariance matrix of the common factors and the diagonal matrix of the latent errors, and it is represented by the following equation: 

\begin{equation}
Cov\left( \mathbf{r}_{t} \right) = \mathbf{\beta}\mathbf{\Sigma}_{\mathbf{f}}\mathbf{\beta}^{T} + \mathbf{D}, 
\end{equation}
where $\mathbf{\Sigma}_{\mathbf{f}}$ is the covariance matrix of the common factors, which can be written as $\mathbf{\Sigma}_{\mathbf{f}}\mathbf{=}\mathbf{f}_{\mathbf{t}}\mathbf{f}_{\mathbf{t}}^{T}$. 

There are two approaches to using factor models. In the first approach, the number of factors and factor scores are assumed to be known, and the factor loadings are computed via regression, as in the CAPM or the Fama-French three- and five-factor models \citep{fama1992cross, fama1993common}. In the second approach, such as in the BARRA factor model (Grinhold and Kahn, 2000), only the number of factors is assumed to be known, and the loadings are estimated, while the factor scores can be computed via regression. If the common factors are assumed to be independent and their scores are appropriately normalized, then the covariance matrix of the common factors is equal to the identity matrix, and it is represented by 
\begin{equation}
\label{Fact_any}
    \mathbf{\Sigma}_{\mathbf{f}}\mathbf{=}\mathbf{I},
\end{equation}
and the covariance of the return series can be expressed as 
\begin{equation}
   \mathbf{\beta}\mathbf{\beta}^{T} + \mathbf{D}.
\end{equation}

By conducting a maximum-likelihood analysis of this equation, policymakers and investors can obtain values for the factor loadings and the latent errors, represented by $\beta_{ij}$ and $\sigma_{i}^{2}$, respectively, where $i = 1,...,N$ and $j = 1,...,m$, given a number of factors $m$.

We conducted an analysis using a portfolio consisting of returns from 10 country indices that were equally weighted. The factor analysis was based on a three-factor model, and Table \ref{Loding} presents the factor loadings and latent error variances estimated from this model. The main question that arises is whether the three factors are sufficient to explain the observed returns. The maximum-likelihood (ML) computation used to produce the data in Table \ref{Loding} assumes that the null hypothesis is that three common factors adequately describe the observed data. The $p$-value statistic obtained from the ML fit to the three-factor model is approximately 0.99, so the null hypothesis is not rejected. However, failing to reject the null hypothesis does not necessarily imply that the null hypothesis is true.

\begin{figure}[h!]
	\centering
	\includegraphics[width=0.75\textwidth]{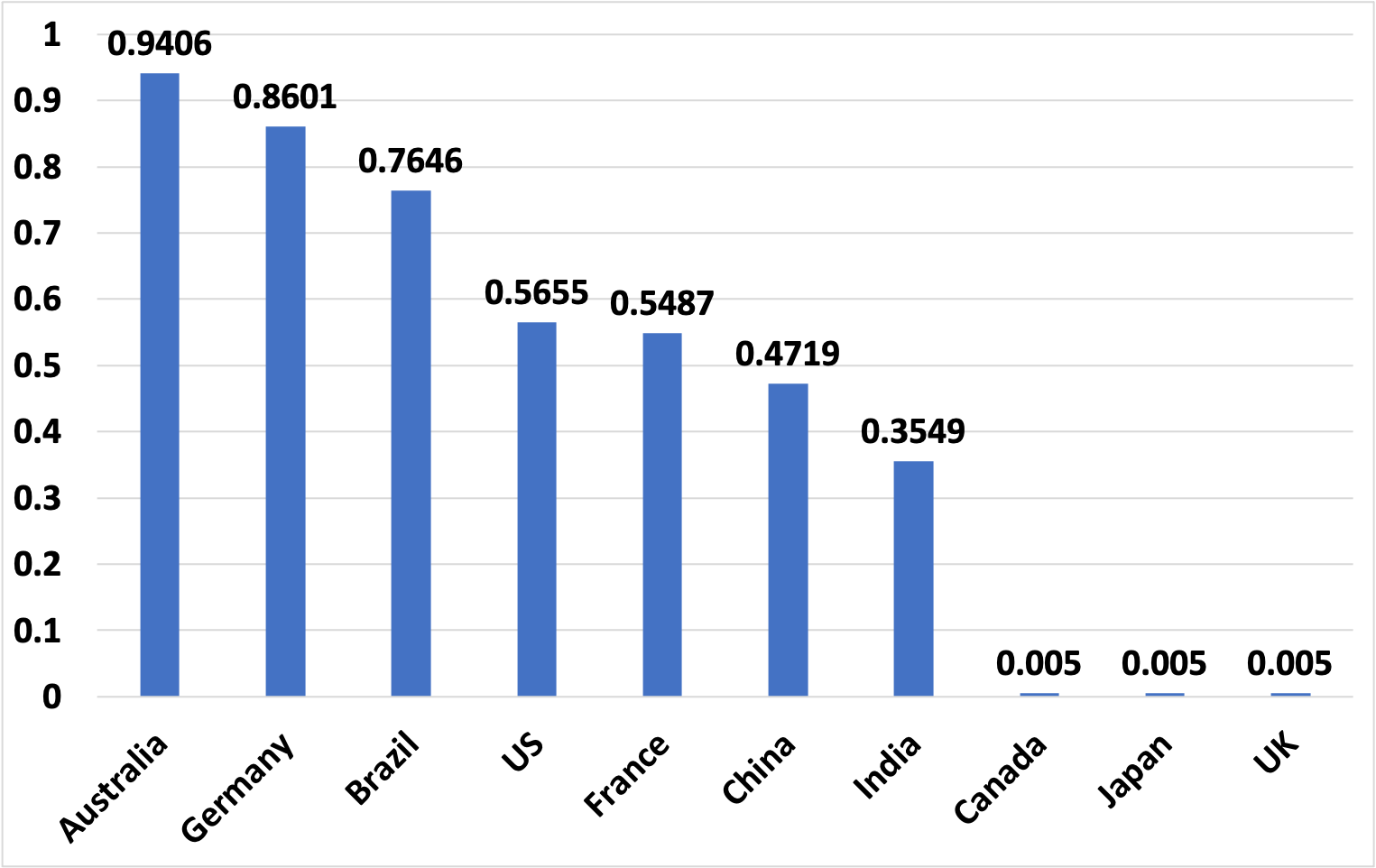} 
	\caption{Indices ordered by latent error variance}
	\label{Fac_load}			
\end{figure}

\begin{figure}[h!]
	\centering
	\includegraphics[width=0.75\textwidth]{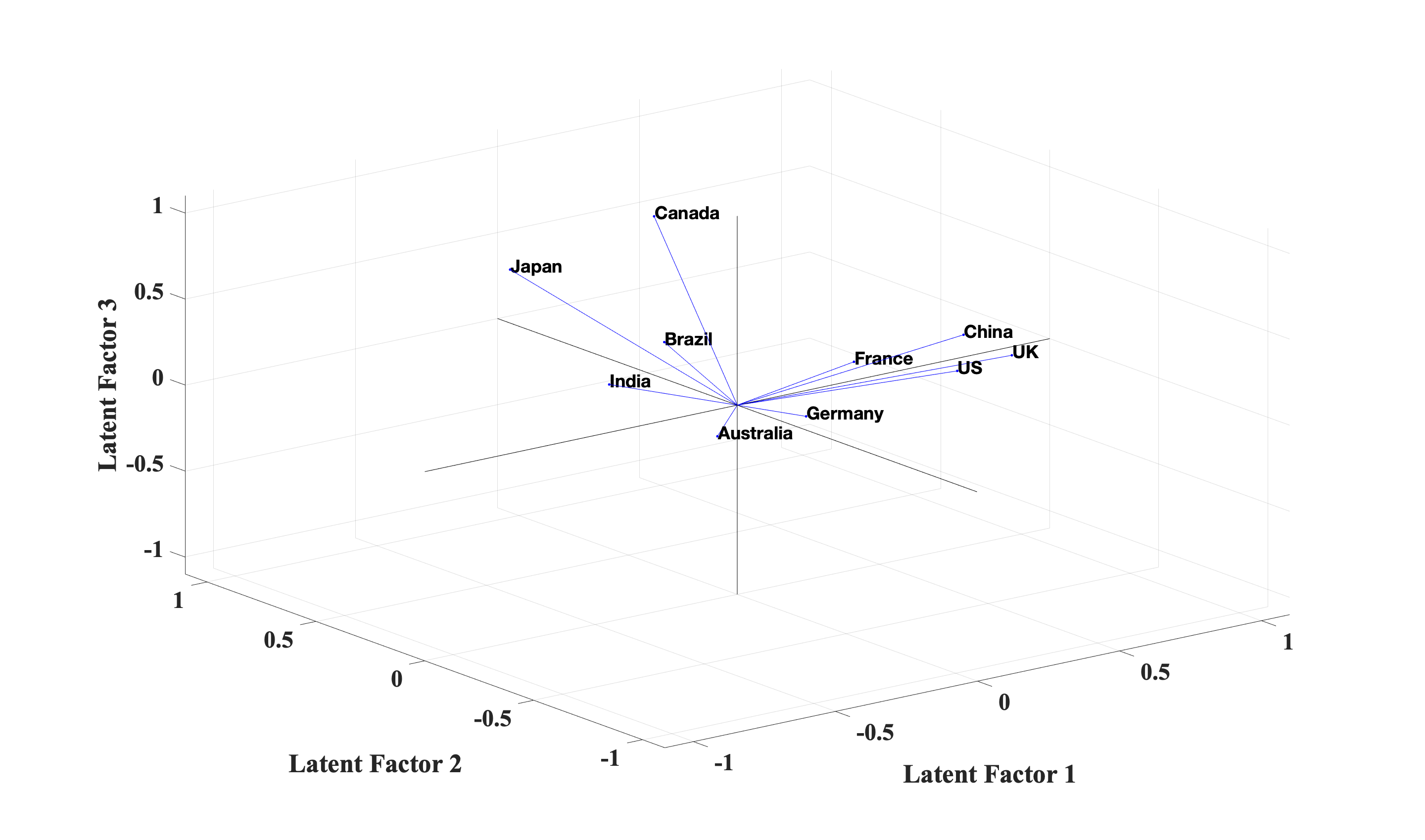} 
	\caption{Loadings of the indices plotted in factor space}
	\label{FA_fig}			
\end{figure}

To further evaluate the adequacy of the three-factor model, we plotted the country indices in Figure \ref{Fac_load} according to their latent error variances. We observe that some country return variances, such as those of Australia, Germany, and Brazil, are not captured well by any of the three common factors. The factor analysis results in Table \ref{Loding} provide an estimate of the variance explained by a model with three common factors. This information offers some insight into the data. To illustrate this insight, we plotted the loading vectors $(\beta_{i1},\beta_{i2},\beta_{i3})$ for each country index in the coordinate space defined by the three factors in Figure \ref{FA_fig}. France, China, the UK, the US, and Germany have strong positive loadings on the first factor, indicating that they share a similar pattern of returns that is captured by this factor. In contrast, Australia is aligned along the positive side of the second factor, indicating that its returns have a different pattern that is not well captured by the other two factors. The countries that have large, positive loadings on the third factor (India, Brazil, Japan, and Canada) appear to share a similar pattern of returns that is not captured well by the other two factors. This suggests that there may be unique economic or market factors driving the returns of these countries that are not captured by the broad market trends represented by the first factor or the country-specific patterns represented by the second factor.

\begin{table}[]
\centering
\begin{tabular}{@{}lcccc@{}}
\toprule
 Country         & $\beta_{i1}$ & $\beta_{i1}$ & $\beta_{i1}$ & $\sigma^2_{i}$ \\ \midrule
Australia & -0.1748      & -0.1489      & -0.0818      & 0.9406         \\
Brazil    & 0.1355       & 0.4476       & 0.1294       & 0.7646         \\
Canada    & 0.1403       & 0.2787       & 0.9474       & 0.005          \\
China     & 0.6          & -0.1423      & 0.3845       & 0.4719         \\
France    & 0.5708       & 0.3541       & -0.0106      & 0.5487         \\
Germany   & 0.2737       & 0.168        & -0.1917      & 0.8601         \\
India     & 0.1214       & 0.7491       & -0.2632      & 0.3549         \\
Japan     & 0.0887       & 0.9399       & 0.322        & 0.005          \\
UK        & 0.9581       & 0.2777       & 0.0021       & 0.005          \\
US        & 0.6536       & 0.0405       & 0.0752       & 0.5655         \\ \bottomrule
\end{tabular}
\caption{Estimated loadings and variance for a three-factor model}\label{Loding}
\end{table}

\newpage

\section{Discussion and Conclusion} \label{sec:DC_CI}
\
\
\
\
\
The development of a financial market for environmental indices has opened up new possibilities for the financial industry to assess, manage, and mitigate potential adverse movements in environmental indicators. This discussion explores the significance of this development and its implications for global environmental conditions and financial risk management.

The goal of creating a global financial market for environmental indices was to allow the financial industry to actively engage with the environments of various countries, including the United States, China, Japan, Germany, the United Kingdom, India, Brazil, Australia, France, and Canada. By incorporating 14 world development socioeconomic indicators, this market provides a comprehensive assessment of environmental well-being.

One key aspect of this market is the evaluation of the risk of downturns in environmental indices worldwide. This evaluation is done using econometric modeling that aligns with dynamic asset pricing theory, allowing for the creation of financial instruments to mitigate these risks. Institutional investors can utilize these instruments to monitor, manage, and trade the indices, thereby creating insurance funds against adverse movements.

While the existing ESG financial markets play an important role in addressing environmental issues, this new global financial market for environmental indices takes a more holistic approach. It encompasses the environmental conditions of countries and provides option prices on these indices, which can serve as insurance instruments against adverse movements. 

A limitation of the current system is the lack of an early warning system for downturns in environmental indices. Gathering more data is necessary to develop an effective system that can predict such downturns for countries worldwide. Additionally, the environmental indices themselves are currently non-tradable, but there is a proposal to introduce ETFs that replicate the dynamics of these indices using fixed-income portfolios.

The potential advantages of this index compared to existing ESG markets lie in its global scope and comprehensive assessment of environmental conditions. While ESG markets primarily focus on socioeconomic issues, the global environmental market covers a broader range of factors. The provision of option prices on the indices as insurance instruments further enhances the utility of this market for risk management.

Future work will involve extending the construction of environmental indices beyond the 10 countries discussed in this paper. The goal is to develop indices for different regions and create a portfolio analysis based on geographical and economic criteria. This expansion would allow for a deeper understanding of the contributions of individual countries or groups of countries to the global environment and their impact.

In conclusion, the establishment of a global financial market for environmental indices provides the financial industry with an opportunity to actively engage in assessing, managing, and mitigating environmental risks. Through the creation of financial instruments and option prices on these indices, institutional investors can monitor and trade them, thereby creating insurance funds against adverse movements. This development represents a significant step towards incorporating environmental factors into financial decision-making and promoting sustainable practices worldwide.

\section{APPENDIX}
\subsection{Environmental factors and economic indicators}

\subsubsection{Adjusted net savings, including particulate emission damage (percentage of GNI)}

 Adjusted net savings are equal to net national savings plus education expenditure minus energy depletion, mineral depletion, net forest depletion, carbon dioxide, and particulate emissions damage.

\subsubsection{Agricultural land (percentage of land area)}

Agricultural land refers to the share of land area that is arable, under permanent crops, and under permanent pastures. Arable land includes land defined by the Food and Agriculture Organization (FAO)\footnote{The Food and Agriculture Organization (FAO) is a United Nations (UN) agency that contributes to international efforts to defeat hunger by developing agriculture. Established in 1945, the FAO is a neutral intergovernmental organization.}\href{https://www.investopedia.com/terms/f/food-agriculture-organization-fao.asp}{} as land under temporary crops (double-cropped areas are counted once), temporary meadows for mowing or for pasture, land under market or kitchen gardens, and land temporarily fallow. Land abandoned as a result of shifting cultivation is excluded. Land under permanent crops is land cultivated with crops that occupy the land for long periods and need not be replanted after each harvest, such as cocoa, coffee, and rubber. This category includes land under flowering shrubs, fruit trees, nut trees, and vines but excludes land under trees grown for wood or timber. Permanent pasture is land used for five or more years for forage, including natural and cultivated crops.

\subsubsection{Arable land (percentage of land area)}

Arable land includes land defined by the FAO as land under temporary crops (double-cropped areas are counted once), temporary meadows for mowing or for pasture, land under market or kitchen gardens, and land temporarily fallow. Land abandoned because of shifting cultivation is excluded.

\subsubsection{CO\textsubscript2 emissions (kt)}

Carbon dioxide emissions are those stemming from the burning of fossil fuels and the manufacture of cement. They include carbon dioxide produced during the consumption of solid, liquid, and gas fuels and gas flaring.

\subsubsection{CO\textsubscript2 emissions (metric tons per capita)}

Carbon dioxide emissions are those stemming from the burning of fossil fuels and the manufacture of cement. They include carbon dioxide produced during the consumption of solid, liquid, and gas fuels and gas flaring.

\subsubsection{The energy intensity level of primary energy}

The energy intensity level of primary energy is the ratio between the energy supply and GDP measured at purchasing power parity. The energy intensity is an indication of how much energy is used to produce one unit of economic output. A lower ratio indicates that less energy is used to produce one unit of output.

\subsubsection{Forest area (percentage of land area)}

The forest area is land under natural or planted stands of trees of at least 5 m on site, whether they are productive or not, and excludes tree stands in agricultural production systems (for example, in fruit plantations and agroforestry systems) and trees in urban parks and gardens.

\subsubsection{Forest area (sq. km)}

The forest area is land under natural or planted stands of trees of at least 5 m on site, whether they are productive or not, and excludes tree stands in agricultural production systems (for example, in fruit plantations and agroforestry systems) and trees in urban parks and gardens.

\subsubsection{Land area (sq. km)}

The land area is a country's total area, excluding areas under inland water bodies, national claims to continental shelves, and exclusive economic zones. In most cases, the definition of inland water bodies includes major rivers and lakes.

\subsubsection{Methane emissions (kt of CO\textsubscript{2} equivalent)}

Methane emissions are those stemming from human activities such as agriculture and industrial methane production.

\subsubsection{Nitrous oxide emissions (thousands of metric tons of CO\textsubscript{2} equivalent)}

Nitrous oxide emissions are emissions from agricultural biomass burning, industrial activities, and livestock management.

\subsubsection{Renewable energy consumption (percentage of total final energy consumption)}

The renewable energy consumption is the share of renewable energy of the total final energy consumption.

\subsubsection{Surface area (sq. km)}

The surface area is a country's total area, including areas under inland bodies of water and some coastal waterways.

\subsubsection{Total natural resources rents (percentage of GDP)}

The total natural resources rents are the sum of oil rents, natural gas rents, coal rents (hard and soft), mineral rents, and forest rents.

\subsubsection{GDP per capita}

The GDP per capita is the GDP divided by the midyear population. The GDP is the total gross value added by all resident producers in the economy plus any product taxes and minus any subsidies not included in the value of the products. It is calculated without making deductions for the depreciation of fabricated assets or for the depletion and degradation of natural resources. The data are in constant 2015 US dollars.

\clearpage

\normalem

\end{document}